\documentclass[reprint,aps,prb,amsmath,amssymb,showpacs,superscriptaddress,footinbib]{revtex4-2}

\usepackage{mathtools}
\usepackage{amssymb}

\usepackage[english]{babel}
\usepackage{csquotes}
\usepackage{dsfont}
\usepackage{xcolor}
\usepackage{amsmath}
\usepackage{hyperref}
\usepackage{booktabs}

\usepackage{graphicx}

\addto\extrasenglish{

}

\newcommand\Eq[1]{\hyperref[#1]{Eq.~(\ref*{#1})}}
\newcommand\eq[1]{\hyperref[#1]{(\ref*{#1})}}

\DeclareMathOperator*\tr{Tr}

\DeclareMathOperator\im{Im}

\usepackage{braket}
\usepackage{simpler-wick}
\renewcommand\Bra[1]{\left(#1\right|}
\renewcommand\Ket[1]{\left|#1\right)}
\newcommand\id{\mathds{1}}

\newcommand\diagram[1]{\,\raisebox{-2ex}{\includegraphics[scale=0.9]{#1.pdf}}\,}
\newcommand\smashdiagram[1]{\,\raisebox{-1ex}{\includegraphics[scale=0.9]{#1.pdf}}\,}

\newcommand\vac{\ensuremath{V_\mathrm{osc}}}
\newcommand\vdc{\ensuremath{V_\mathrm{avg}}}
\newcommand\gac{\ensuremath{G_\mathrm{osc}}}
\newcommand\gdc{\ensuremath{G_\mathrm{avg}}}
\newcommand\iac{\ensuremath{I_\mathrm{osc}}}
\newcommand\idc{\ensuremath{I_\mathrm{avg}}}
\newcommand\tkv{\ensuremath{T_K}}
\newcommand\tkrg{\ensuremath{T_K'}}

\newcommand\Ra{R}
\newcommand\Leff{\hat{L}}
\newcommand\Gfull{G}
\newcommand\Gkavg{\breve{G}}
\newcommand\Gnoeta{\bar{G}}
\newcommand\Ifull{I}

\newcommand\Inoeta{\bar{I}} 

\begin{document}
\title{
Floquet Renormalization Group Approach to the Periodically Driven Kondo Model
}

\author{Valentin Bruch}
\affiliation{Institut f\"ur Theorie der Statistischen Physik, RWTH Aachen,
52056 Aachen, Germany and JARA--FIT, 52056 Aachen, Germany}

\author{Mikhail Pletyukhov}
\affiliation{Institut f\"ur Theorie der Statistischen Physik, RWTH Aachen,
52056 Aachen, Germany and JARA--FIT, 52056 Aachen, Germany}

\author{Herbert Schoeller}
\affiliation{Institut f\"ur Theorie der Statistischen Physik, RWTH Aachen,
52056 Aachen, Germany and JARA--FIT, 52056 Aachen, Germany}

\author{Dante M. Kennes}
\affiliation{Institut f\"ur Theorie der Statistischen Physik, RWTH Aachen,
52056 Aachen, Germany and JARA--FIT, 52056 Aachen, Germany}
\affiliation{Max Planck Institute for the Structure and Dynamics of Matter, Center for Free Electron Laser Science, Luruper Chaussee 149, 22761 Hamburg, Germany}
\email{Dante.Kennes@rwth-aachen.de}

\pacs{
}

\begin{abstract}
We study the interplay of strong correlations and coherent driving by considering the strong-coupling Kondo model driven by a time-periodic bias voltage. Combining a recent nonequilibrium renormalization group method with Floquet theory, we find that by the coherent dressing of the driving field side-replicas of the Kondo resonance emerge in the conductance, which are not completely washed out by the decoherence induced by the driving. We show that to accurately capture the interplay of driving and strong correlations one needs to go beyond simple phenomenological pictures, which underestimate decoherence, or adiabatic approximations, highlighting the relevance of non-Markovian memory effects.
Within our method the differential conductance shows good quantitative agreement with experimental data in the full crossover regime from weak to strong driving.
We analyze memory effects in detail based on the response to short voltage pulses.
\end{abstract}

\maketitle

\section{Introduction}
\begin{figure}
  \centering
\includegraphics[width=.9\linewidth]{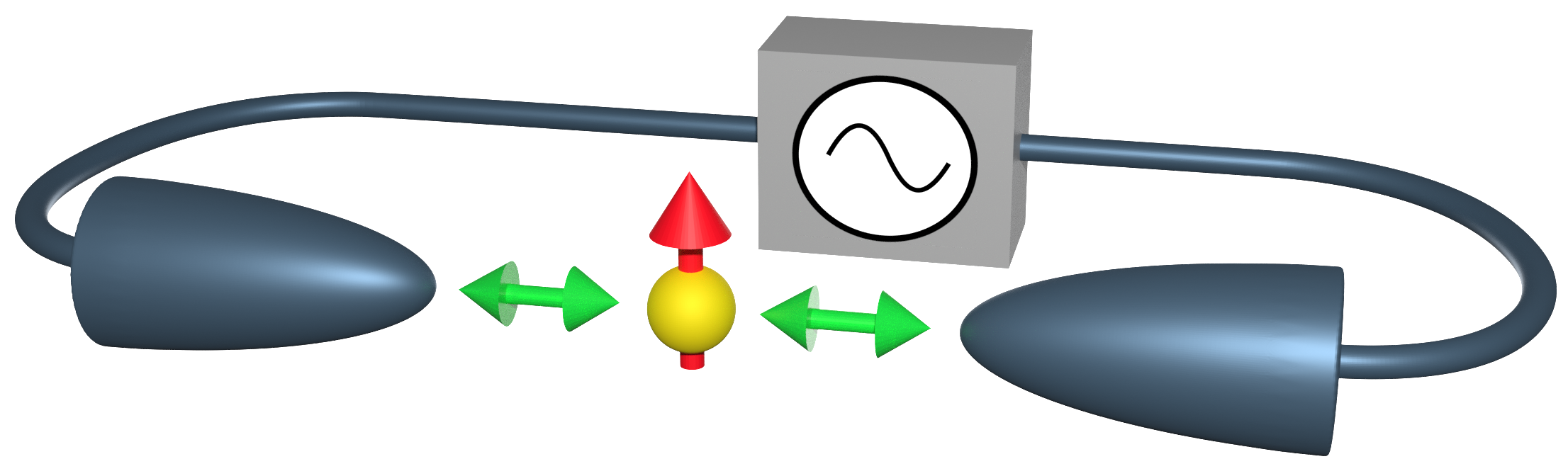}
  \caption{\label{fig:model}Sketch of the model. A periodic bias voltage is applied across a quantum dot, which in the Kondo limit can be reduced to a single spin-$\frac12$ (red) coupled to two fermionic reservoirs $L$ and $R$ (blue).
  }
\end{figure}

The nonequilibrium Kondo model has drawn interest for many years because it highlights quantum many-body effects in a basic, yet physically relevant and experimentally accessible model~\cite{Hewson93}.
In its prototypical form a single spin-$\tfrac{1}{2}$ is coupled to the spins of two fermionic reservoirs at low temperature as sketched in \autoref{fig:model}.
This model can be realized in a highly controllable manner by a quantum dot in the Coulomb blockade regime coupled to two reservoirs, with a controllable coupling and bias voltage \cite{Glazman88,Ng88,Goldhaber-Gordon98,Cronenwett98,Simmel99}.
In this realization the Kondo effect causes an enhanced differential conductance for vanishing bias voltage in the low temperature limit.
Although the equilibrium properties of this model are by now well understood \cite{Andrei80,Ludwig91,Affleck93}, the suppression of the Kondo resonance in nonequilibrium conditions and the dynamics of the model remain topics of recent research \cite{Nghiem17,Bruhat18,Kanasz-Nagy18,Krivenko19}.

The low-energy physics of the Kondo model in equilibrium is controlled by a single characteristic energy scale, the Kondo temperature $\tkv$.
In nonequilibrium one distinguishes between the strong coupling case, for which the Kondo temperature is large compared to the bias voltage and temperature, and the opposite case of weak coupling, when either the temperature or bias voltage is large compared to $\tkv$.
Various approaches have been used to describe the Kondo model in both regimes, and different perturbative renormalization group (RG) methods have been applied successfully for weak and intermediate coupling \cite{Wilson75,Kaminski99,Kaminski00,Rosch01,Rosch03,Kehrein05,Doyon06,Korb07,Schoeller09b,Pletyukhov10,Pletyukhov12a,Reininghaus14,Horig14}.
However, most of these works consider a constant bias voltage. Here, we focus on a general time-dependent, periodic bias voltage.
Periodic driving has been discussed in the form of an oscillating gate voltage applied to the quantum dot that effectively leads to a time-dependent coupling \cite{Hettler95,Lopez98,Kaminski99,Kaminski00}, and an oscillating bias voltage \cite{Ng96,Goldin98,Kaminski00} as considered here.
In the latter case the differential conductance has been analyzed in the case of fast driving ($\hbar\Omega\gg \max\{e\vac,\tkv\}$) using a renormalization group technique, and in the opposite limit of slow driving ($\hbar\Omega\ll\max\{e\vac,\tkv\}$) using the adiabatic limit \cite{Kaminski00}.
Here $\Omega$ is the driving frequency and $\vac$ is the driving amplitude.
The crossover between slow and fast driving ($\hbar\Omega\sim e\vac\gg\tkv$) has been approximated using bare perturbation theory \cite{Kaminski00}.
But although the results of Ref.~\cite{Kaminski00} provide a valuable overview with predictions over different parameter regimes, it does not include some resonance effects predicted by the results of Ref.~\cite{Goldin98} using a perturbation expansion in the system-reservoir coupling.
Other works discussing this setup were based on a noncrossing approximation~\cite{Hettler95}, an equation of motion approach~\cite{Ng96}, or an interpolated effective self-energy~\cite{Lopez98},
all of which involve simplifying assumptions regarding decoherence due to the oscillating voltage.
In this paper we complement these approaches by focusing on the questions to what extent quantum effects are suppressed by an alternating bias voltage and which effects dominate the micromotion of the system within a period of driving.

It has been proposed that alternating fields should split the Kondo resonance in the density of states to form photon sidebands~\cite{Hettler95,Goldin98} as one expects for usual photon-assisted tunneling~\cite{Tien63}.
This effect leads to satellite peaks in the differential conductance, which have been observed in quantum dots coupled to microwave radiation~\cite{Kogan04,Bruhat18}.
Indications of the same effect have also been found in a single molecule transistor~\cite{Yoshida15}.
Other experiments studying quantum dots in the presence of microwave radiation could not see indications of photon sidebands presumably due to noise or due to a low driving frequency~\cite{Elzerman00}.
Similar side peaks in the differential conductance have been predicted~\cite{Paaske05} and observed~\cite{Yu04,Yu04b,Parks07} when a quantum dot in the Kondo regime is coupled to a vibrational mode.
However, the precise role of decoherence in the suppression of the central Kondo resonance has remained unclear.

In this work we extend the real-time renormalization group (RTRG) technique of Refs.~\cite{Schoeller09,Pletyukhov12a,Kashuba13a,Reininghaus14,Horig14,Lindner19} to the case of periodic driving by employing Floquet theory in Liouville space~\cite{Tsuji08,Mathey20}.
The RTRG has been used to describe the stationary limit of the Kondo model at constant bias voltage in good agreement with other methods and experiments~\cite{Pletyukhov12a,Reininghaus14,Kretinin11,Kretinin12}.
Extending it to what we will call Floquet RTRG (FRTRG) allows us to describe the coherently driven Kondo model for an arbitrary periodic driving profile.

This paper is organized as follows: Section \ref{sec:model} briefly introduces the model.
Sections \ref{sec:basics}--\ref{sec:initial-conditions} contain an extensive derivation of the method that can be skipped by readers who are only interested in the results.
In \autoref{sec:basics} the basic notation for open system evolution in Liouville space combined with Floquet theory is introduced, including a diagrammatic language that simplifies a formal expansion in the coupling of the system to the reservoirs.
This diagrammatic language is used in \autoref{sec:general-RG} to derive RG equations on a general level without using the model-specific algebraic structure in Liouville space to avoid unnecessary complications.
\autoref{sec:kondo-RG} contains the technical and model-specific details of the derivation of the RG equations for the Kondo model.
The initial conditions required to solve these RG equations are subsequently derived in \autoref{sec:initial-conditions}.
\autoref{sec:method-summary} provides a brief summary of the method focused on the practical computation of observables and introduces some notation for \autoref{sec:results}.
In \autoref{sec:results} the results are discussed and compared to experimental data.
We summarize the most important results and ideas in \autoref{sec:summary}.

\section{Model}
\label{sec:model}

We study the prototypical spin-$\tfrac{1}{2}$ Kondo model described by the Hamiltonian~\cite{Glazman05}
\begin{equation}
  H(t) = \sum_{\alpha\alpha'} J^{(0)}_{\alpha\alpha'} \mathbf{S}\cdot\mathbf{s}_{\alpha\alpha'} + \sum_{\alpha\sigma} \int d\omega\, [\omega+\mu_\alpha(t)] c_{\alpha\sigma\omega}^\dag c_{\alpha\sigma\omega} ,
\end{equation}
where $\mathbf{S}$ denotes the impurity spin which is isotropically coupled to the spins $\mathbf{s}_{\alpha\alpha'} = \tfrac12 \sum_{\sigma\sigma'}\iint d\omega d\omega'\, \boldsymbol{\sigma}_{\sigma\sigma'} c_{\alpha\sigma\omega}^\dag c_{\alpha'\sigma'\omega'}$
of two reservoirs denoted by left ($\alpha=L$) and right ($\alpha=R$). 
We define the bare coupling $J^{(0)}_{\alpha\alpha'}=2\sqrt{x_\alpha x_{\alpha'}} J^{(0)}$, chemical potentials $\mu_\alpha(t)$ and electron annihilation (creation) operators $c_{\alpha\sigma\omega}^{(\dag)}$ of a state with energy $\omega+\mu_\alpha(t)$ and spin $\sigma$ in reservoir $\alpha$. $\boldsymbol{\sigma}_{\sigma\sigma'}$ denotes the vector of Pauli matrices and $x_L=1-x_R$ characterizes the asymmetry of the coupling to the two reservoirs.
To drive the system out of its equilibrium state we consider a periodic bias voltage $V(t)=\mu_L(t) - \mu_R(t)$ across the reservoirs and mainly focus on the case $V(t)=\vdc+\vac\cos(\Omega t)$. However, we stress that the machinery developed here can be applied to any form of time-periodic driving. We use units $e=\hbar=k_B=1$.

The reservoirs are assumed to be non-interacting and have a density of states of the form $\varrho_{\alpha\sigma}(\omega) = \varrho_{\alpha\sigma}^{(0)} D(\omega)$ where $\varrho_{\alpha\sigma}^{(0)}$ is absorbed in the definition of the creation and annihilation operators and the cutoff function $D(\omega)=D^2/(D^2+\omega^2)$ remains in the anticommutation relation of these operators:
\begin{equation}
  \{ c_{\alpha\sigma\omega}, c^\dag_{\alpha'\sigma'\omega'} \} = D(\omega) \delta(\omega-\omega') \delta_{\sigma\sigma'} \delta_{\alpha\alpha'}
  .
\end{equation}
To study the universal low-energy physics we take the limit of an infinite band width $D\to\infty$ leading to structureless wideband reservoirs, while adjusting the bare coupling $J^{(0)}$ such that the Kondo temperature $\tkv$ remains finite~\cite{Reininghaus14}.
Throughout the paper we work with reservoirs at zero temperature, which significantly reduces the analytical and numerical effort (see Ref.~\cite{Reininghaus14} for a finite-temperature RTRG calculation).

\section{Liouville-Floquet space and diagrammatic language}
\label{sec:basics}

To apply the RTRG to periodically driven systems we combine the RTRG with Floquet theory in Liouville space \cite{Tsuji08,Mathey20}.
In this section we explain the notation used in combined Floquet-Liouville space including the diagrammatic language that will be used to derive the RG equations.
An overview of the notation used in the derivation of the method is provided in \autoref{tab:notation} in App.~\ref{app:notation}.

\subsection{Floquet theory in Liouville space}
\label{sec:floquet}
Floquet theory in Liouville space allows us to benefit from Fourier transforms not only in the time-translation invariant but also in the periodically driven case and has been developed within the Floquet Green's function formalism \cite{Tsuji08,Wu08,Wu10,Genske15,Eissing16,Eissing16b,Mathey20}.
In the Floquet formalism we consider observables in periodically driven systems that depend on two time arguments. Typically these are the initial time $t_0$ when the evolution started, and the time $t$ at which the observable is measured.
To introduce the formalism we consider a general function $f(t,t')$ of two time arguments $t\geq t'$.
Due to the periodic driving, the whole system is invariant under a shift of all time variables by one time period $T$, $f(t,t')=f(t+T,t'+T)$.
A Fourier transform in the relative time $t-t'$ and a Fourier series expansion in $t$ yield~\cite{Stefanucci08,Tsuji08}
\begin{equation}
\tilde{f}_n(E) \coloneqq \frac{1}{T} \int_0^\infty ds\, e^{iEs} \int_0^T dt\, e^{in\Omega t} f(t, t-s) \,.
  \label{eq:floquet_1}
\end{equation}
Here the sign of $E$ in the Fourier transform is chosen such that $\tilde{f}_n(E)$ is analytic in the upper half of the complex plane.

A Floquet matrix can now be defined by \cite{Tsuji08,Wu08,Wu10,Genske15,Mathey20}
\begin{equation}
  \hat{f}(E)_{nm}\coloneqq\tilde{f}_{n-m}(E+m\Omega) \,.
  \label{eq:floquet-matrix}
\end{equation}
This definition has the advantage that convolutions in time domain become matrix products in Floquet space (cf.~App.~\ref{app:floquet}).
We will see below that this property leads to a close analogy between the RTRG for time-translation invariant systems and the Floquet RTRG.

It is instructive to consider the case of vanishing external driving as a special case of the Floquet formalism.
In this limit of a time-translation invariant system a function of two time arguments $f(t,t')$ can only depend on the relative time $t-t'$ and not on the absolute time $t$.
In this case the Floquet matrix $\hat{f}(E)$ of the function $f(t,t')=f(t-t')$ becomes $\hat{f}(E)_{nm}=\delta_{nm}f(E+n\Omega)$ where $f(E)=\int_0^\infty ds\,e^{iEs}f(s)$ is the Fourier transform of $f(t-t')$.
In matrix notation we write this as
\begin{equation}
  \hat{f}(E) = f(E+\hat{N}\Omega)
  \,,\qquad
  \hat{N}_{nm} = n\delta_{nm}\,,
  \label{eq:fourier-floquet}
\end{equation}
where $\hat{N}$ is a diagonal Floquet matrix.
Since in this limiting case all Floquet matrices are diagonal, it is sufficient to consider only the matrix element $\hat{f}(E)_{00}$ which is just the Fourier transform of $f(t-t')$ reproducing the well-known case of an undriven system.

\subsection{Evolution in Floquet-Liouville space}
The derivation of the RG equations is based on an expansion of the evolution in the coupling between system and reservoirs. Here we introduce the notation to describe this expansion in Liouville space for the Kondo model in close analogy to Ref.~\cite{Reininghaus14}.
The dynamical map $\Pi(t,t_0)$ describing the evolution of the density matrix of the central spin from time $t_0$ to $t$ by $\Pi(t,t_0)\rho(t_0)=\rho(t)$ can formally be written in the form
\begin{equation}
  \Pi(t,t_0) = \tr_R \Big\{ \mathcal{T} e^{-i\int_{t_0}^t ds\,\big[ L_V + L_R(s) \big]} (\bullet\otimes\rho_R^\mathrm{eq}) \Big\} \,,
  \label{eq:pi-time-domain}
\end{equation}
assuming that at time $t_0$ all reservoirs are separately in equilibrium (state $\rho_R^\mathrm{eq}$).
The dynamical map is a superoperator, \emph{i.e.} a linear map acting on operators like density matrices.
Here $\bullet$ denotes the operator on which $\Pi(t,t_0)$ acts, $\tr_R$ is the trace over the reservoir degrees of freedom and $\mathcal{T}$ denotes time ordering.
We furthermore introduced the time-periodic reservoir Liouvillian
\begin{equation}
  L_R(t) = \sum_\alpha \int d\omega [\omega+\mu_\alpha(t)] \sum_\sigma [c_{\alpha\sigma\omega}^\dag c_{\alpha\sigma\omega}, \bullet]
  \label{eq:reservoir-liouvillian}
\end{equation}
and a coupling Liouvillian $L_V$.
For now we can avoid complications by working with a general Liouvillian $L_V$ that acts on density matrices of the total system (including the reservoirs).
As derived in App.~\ref{app:floquet}, in Floquet space the dynamical map takes the form
\begin{align}
  \hat{\Pi}(E) &= \tr_R \Big\{ \frac{i}{E+\hat{N}\Omega - \hat{L}_R - L_V} (\bullet\otimes\rho_R^\mathrm{eq}) \Big\}
  \label{eq:pi-floquet-space} \\
  &= \frac{i}{E+\hat{N}\Omega - \Leff(E)} \,, \label{eq:L-floquet-space} \\
  \hat{L}_{R\,nm} &= \frac1T \int_0^T dt\, e^{i(n-m)\Omega t} L_R(t) \,,
  \label{eq:L_R-floquet-space}
\end{align}
where we marked all Floquet matrices with a hat and introduced the effective Liouvillian $\Leff(E)$.
The coupling Liouvillian $L_V$ does not become a Floquet matrix because it is time-independent.
In the diagrammatic language which we will define below, all possible diagrams will form $\hat\Pi(E)$ and a subset of (irreducible) diagrams will sum up to $\Leff(E)$.

Before we continue towards evaluating \Eq{eq:pi-floquet-space} in close analogy to Ref.~\cite{Reininghaus14} we need to introduce a compact notation.
To avoid working with many indices we collect multiple indices like $\alpha,\sigma,\omega$ in multi-indices which we denote by digits.
The multi-index $1\equiv(\alpha_1,\sigma_1,\omega_1,\eta_1)$ includes the reservoir, spin and energy indices, and additionally distinguishes between fermion creation $(\eta=+)$ and annihilation $(\eta=-)$ operators. In cases where we need only two of these multi-indices we use the simplified notation $1=(\alpha,\sigma,\omega,\eta)$ and $1'=(\alpha',\sigma',\omega',\eta')$.
By $\bar1$ we denote the same multi-index with $\eta$ flipped to $\bar\eta=-\eta$.
For example, if $\eta=+$ we can write $c_1=c^\dag_{\alpha\omega\sigma}$ and $c_{\bar1}=c_1^\dag$.
With this notation we can now define fermionic superoperators,
\begin{equation}
  J^+_1 = c_1 \bullet\,,\qquad J^-_1 = \bullet c_1\,,
  \label{eq:fermionic-superoperators}
\end{equation}
by left or right multiplication with a creation or annihilation operator~\cite{Schoeller09}.
The upper index $\pm$ of these superoperators corresponds to the Keldysh index which distinguishes between forward ($+$) and backward ($-$) propagation in the usual Keldysh formalism. Although we do not use this formalism here, we still refer to this index as the Keldysh index.

Focusing on the calculation of $\hat{\Pi}(E)$ using \Eq{eq:pi-floquet-space} again, we need a formalism to trace out the reservoirs.
We will do so by expanding in $L_V$ and using Wick's theorem, but this requires that we first write $L_V$ in a form that uses fermionic superoperators in Liouville space.
In the Kondo model the coupling Hamiltonian connecting system and reservoirs is of the form
\begin{equation}
  V = \frac12 \sum_{\alpha\alpha'\sigma\sigma'} J^{(0)}_{\alpha\alpha'} \mathbf{S}\cdot\boldsymbol{\sigma}_{\sigma\sigma'} \iint d\omega d\omega'\,c_{\alpha\sigma\omega}^\dag c_{\alpha'\sigma'\omega'}\,,
  \label{eq:coupling-hamiltonian}
\end{equation}
which is bilinear in the reservoir creation and annihilation operators.
The Liouvillian $L_V=[V,\bullet]$ for any operator $V$ that is bilinear in the fermionic operators in the reservoirs can be written in the general form
\begin{equation}
L_V = \frac12 p' \Gfull^{(0)pp'}_{11'} :J_1^p J_{1'}^{p'}:
  \label{eq:vertex-liouvillian}
\end{equation}
that defines the bare coupling vertex $\Gfull^{(0)pp'}_{11'}$ which acts only on the system and not on the reservoirs.
Here $:\ldots:$ denotes normal ordering of the reservoir field operators and summation over equal (multi-)indices $p,p',1,1'$ is implicit.
The explicit expression for the bare coupling vertex will be provided below [\Eq{eq:bare-vertex}], but for now it will be simpler to consider a frequency-independent but otherwise general superoperator $\Gfull^{(0)pp'}_{11'}$.
To avoid the ambiguity in the definition of the bare coupling vertex in \Eq{eq:vertex-liouvillian} we furthermore require that $\Gfull^{(0)p'p}_{1'1} = - \Gfull^{(0)pp'}_{11'}$
\footnote{A possible component of $\Gfull^{(0)pp'}_{11'}$ that is symmetric under exchanging $1$ with $1'$ and $p$ with $p'$ will not contribute to \Eq{eq:vertex-liouvillian} because $:J^p_1 J^{p'}_{1'}:=-pp':J_{1'}^{p'} J^p_1:$.}.

In order to trace out the reservoirs in \Eq{eq:pi-floquet-space}
we further need the commutation relation of the fermion superoperators with the reservoir Liouvillian,
\begin{equation}
  J_1^\pm \hat{L}_R = (\hat{L}_R-\eta\omega+\eta\hat{\mu}_\alpha) J_1^\pm \,.
\end{equation}
Here the Floquet matrix $\hat\mu_\alpha$ is defined analogous to \Eq{eq:L_R-floquet-space} based on $\mu_\alpha(t)$.
This enables us to calculate the leading order contribution to the effective Liouvillian,
\begin{align}
  \hat{L}^{(2)}(E) &= \tr_R \Big\{ L_V \frac 1{E+\hat{N}\Omega-\hat{L}_R} L_V (\bullet \otimes \rho_R^\mathrm{eq}) \Big\} \\
  &= \frac14 p_2 \Gfull^{(0)p_1p_2}_{12} \hat{R}^{(0)}_{12}(E) p_4 \Gfull^{(0)p_3p_4}_{34} \times\notag \\
  &\qquad \times \tr_R (:J_1^{p_1} J_2^{p_2}:\,:J_3^{p_3} J_4^{p_4}:\rho_R^\mathrm{eq}) \,,
  \label{eq:L-leading-order-Floquet-space}
\end{align}
where we defined the bare resolvent
\begin{equation}
  \hat{R}^{(0)}_{12}(E) = \frac{1}{E + \hat{N}\Omega - \hat{\bar\mu}_{12} + \bar\omega_{12}}
  \label{eq:bare-resolvent}
\end{equation}
with the notation $\bar\omega_{12\ldots} = \eta_1\omega_1 + \eta_2\omega_2 + \ldots$ and analogous $\hat{\bar\mu}_{1\ldots}=\eta_1\hat\mu_1+\ldots$.
Summation over \mbox{(multi-)}indices $1,p_1,\ldots$ and integration over frequencies are implicit.
The reservoir contribution in \Eq{eq:L-leading-order-Floquet-space} can be evaluated using Wick's theorem for the superoperators $J_n^p$~\cite{Schoeller09} by summing up all possible contractions of the field superoperators $:J_1^{p_1}J_2^{p_2}:$ with $:J_3^{p_3}J_4^{p_4}:$,
\begin{multline}
  \tr_R (p_2:J_1^{p_1} J_2^{p_2}:\,p_4:J_3^{p_3} J_4^{p_4}:\rho_R^\mathrm{eq}) =\\
  \gamma^{p_1p_4}_{14}(\omega_1,\omega_4)\gamma^{p_2p_3}_{23}(\omega_2,\omega_3) -
  \gamma^{p_1p_3}_{13}(\omega_1,\omega_3)\gamma^{p_2p_4}_{24}(\omega_2,\omega_4) \,,
  \label{eq:wick}
\end{multline}
with the reservoir contraction function
\begin{align}
  \gamma^{pp'}_{11'}(\omega,\omega') &= \delta_{1\bar{1}'}\delta(\eta\omega + \eta'\omega') \gamma^{p'}(\eta\omega), \\
  \gamma^{p}(\omega) &= p \frac1{1+e^{p\omega\beta}} D(\omega)
  \label{eq:reservoir-contraction}
\end{align}
at temperature~$1/\beta$. Here we wrote the frequency dependence explicitly that is usually contained in the multi-indices.
The general form of Wick's theorem for the superoperators $J_1^p$ and its derivation can be found in App.~\ref{app:wick} and in Sec.~3.1 of Ref.~\cite{Schoeller09}.
The minus sign in \Eq{eq:wick} results from the odd number of permutations of the indices $1,2,3,4$ in the second term.
Here we are only interested in the low temperature limit,
\begin{equation}
  \gamma^p(\omega) = p \Theta(-p\omega) D(\omega) \,,
  \label{eq:reservoir-contraction-T0}
\end{equation}
where $\Theta$ denotes the step function.

By inserting \Eq{eq:wick} in \Eq{eq:L-leading-order-Floquet-space} and using the symmetry $\Gfull_{11'}^{(0)pp'}=-\Gfull_{1'1}^{(0)p'p}$ we can write $L^{(2)}(E)$ in the form
\begin{equation}
  \hat{L}^{(2)}(E) = \frac12 \Gfull^{(0)p_1p_2}_{12} \hat{R}^{(0)}_{12}(E) \Gfull^{(0)p_3p_4}_{34}\, \gamma^{p_1p_4}_{14} \gamma^{p_2p_3}_{23} \,.
  \label{eq:L-leading-order-2}
\end{equation}
Next we will define a diagrammatic language for such expressions that consist only of the building blocks $\Gfull^{(0)}$, $\hat{R}^{(0)}(E)$, and $\gamma$.

\subsection{Diagrammatic language}
The right hand side of \Eq{eq:L-leading-order-2} can be written as a diagram
consisting of bare vertices~$\Gfull^{(0)}$ (double circles) that are connected by bare resolvents~$\hat{R}^{(0)}$ (black lines) and by reservoir contractions~$\gamma$ (green lines):
\begin{align}
  &\Leff^{(2)}(E) = \diagram{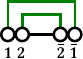} \label{eq:example-diagram}\\
&= \frac1{2!} \gamma^{p_4}(\bar\omega_1) \gamma^{p_3}(\bar\omega_2) \Gkavg_{12}^{(0)} \hat{R}^{(0)}_{12}(E) \Gfull_{\bar{2}\bar{1}}^{(0)p_3p_4} \,.
  \label{eq:L-leading-order-simplified}
\end{align}
Here the bare vertex $\Gkavg_{12}^{(0)}$ without Keldysh indices denotes $\sum_{pp'} \Gfull^{(0)pp'}_{11'}$ and again integration over frequencies is implicit.

Also all other contributions to the expansion of $\hat{\Pi}(E)$ in $L_V$ can be written in this diagrammatic language and the sum of all possible diagrams yields precisely $\hat{\Pi}(E)$. By comparison with an expansion of \Eq{eq:L-floquet-space} one can find the subset of diagrams that form $\Leff(E)$. The effective Liouvillian is given by the sum of all diagrams that remain connected by contraction lines when cutting an arbitrary reservoir propagation line.

The rules for translating diagrams can be summarized as follows.
A contraction line connects a multi-index $1$ with $\bar1$ and represents the scalar function $\gamma^{p}(\bar\omega_1)$. Each crossing of two contraction lines contributes a minus sign.
The other components of the diagram represent superoperators and Floquet matrices which generally do not commute.
A horizontal black line represents a resolvent $\hat{R}^{(0)}_X(E)$ [\Eq{eq:bare-resolvent}] where $X$ denotes the set of indices of the contraction lines above this resolvent.
By convention, the index of a contraction line $\gamma^p(\bar\omega_1)$ connecting indices $1$ and $\bar{1}$ is its left index ($1$).
The diagram shown in Eq.~\eq{eq:example-diagram} furthermore includes a symmetry factor $1/S$ where $S=\prod_k m_k!=2!$ collects for each pair of vertices the number $m_k$ of contraction lines by which these vertices are directly connected~\cite{Schoeller09}.
The effective Liouvillian $\Leff(E)$ is given by the sum of all irreducible diagrams ({\it i.e.} diagrams connected by reservoir lines) with no free reservoir lines.
Diagrams which only differ by the order in which contraction lines are connected to the same vertex are equivalent and should be counted only once.

The strength of the RTRG is that it includes a large class of diagrams by self-consistently replacing the bare components of the diagrams with effective vertices and resolvents.
The first step is to replace the bare resolvent $\hat{R}^{(0)}(E)$ by an effective one including all self-energy insertions:
\begin{align}
  \hat{R}(E) &= \frac1{E + \hat{N}\Omega - \Leff(E)} \,, \label{eq:effective-resolvent} \\
  \hat{R}_X(E) &= \hat{R}(E-\hat{\bar\mu}_X+\bar{\omega}_X) \,. \label{eq:effective-resolvent-argument}
\end{align}
From here on these effective resolvents will be used instead of the bare ones in all diagrams. This implies that a diagram which differs from another diagram only by a self-energy insertion, \emph{i.e.} by an insertion of a diagram for $\Leff(E)$ in a resolvent line, should not be counted as a new diagram.

Also the vertices can be renormalized.
A renormalized $n$-point vertex consists of all irreducible diagrams with $n$ open contraction lines. In the present paper we will only need the two-point vertex or effective coupling vertex, but four-point vertices would be required to include higher order corrections to our calculation.
From this definition of the effective vertex we can directly see that all diagrams for $\Leff(E)$ can be obtained by connecting an effective vertex and a bare vertex.
Thus a naive replacement of bare vertices with renormalized ones would lead to double counting of some diagrams. In the RTRG this problem is avoided by splitting diagrams with an energy derivative as we will see below.
Here we introduce the rules for diagrams with free reservoir lines that define the effective vertex.
As an example diagram we consider the leading order correction to the bare vertex,
\begin{align}
  \label{eq:effective_vertex}
  &\Gfull^{p_1p_2}_{12}(E;\bar\omega_1,\bar\omega_2) = \Gfull^{(0)p_1p_2}_{12} \notag \\
  &\qquad\qquad + \diagram{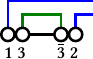} + \diagram{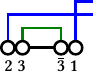} + O(G^3) \,,\\
  &\diagram{G2example1} = \gamma^{p_4}(\bar\omega_3) \Gfull^{(0)p_1p_3}_{13} \hat{R}_{13}(E) \Gfull^{(0)p_4 p_2}_{\bar{3}2} \,.
\end{align}
We write the frequency dependence of the effective vertex explicitly to stress the difference to the frequency-independent bare vertex.
When computing these diagrams all free reservoir lines are by convention directed to the right to ensure the correct energy shift in the propagator $\hat{R}_{13}(E)$.

In the construction of the effective vertex one has to sum over all permutations of external lines. As an exception, external lines that are connected to the same bare vertex should not be permuted.
Eventually, only the effective vertex at vanishing external frequencies, $\omega_1=\omega_2=0$, will be required in the RG equations.
The diagrams remind us that since $\hat{R}(E)$ is a Floquet matrix, also the effective Liouvillian $\Leff(E)$ and the effective vertex $\Gfull^{pp'}_{11'}(E;\omega,\omega')$ will be Floquet matrices.

\section{General RG equations}
\label{sec:general-RG}

Our aim when deriving the RG equations is to find a closed set of differential equations describing the $E$-dependence of the effective Liouvillian $L(E)$. We will then extend the equations to include observables, most notably the current.
From here on Floquet matrices will not be marked with a hat anymore since most of the objects appearing in these calculations will be Floquet matrices.
The derivation closely follows Ref.~\cite{Reininghaus14} that considered a time-translation invariant system.

\subsection{Energy derivatives}

We recall that the leading order diagram for $L(E)$, $\diagram{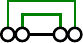}$ contains two frequency integrals and one resolvent [\Eq{eq:L-leading-order-simplified}]. This diagram is not convergent in the limit $D\to\infty$, but can be regularized by taking the second derivative of the resolvent.
The energy-dependence of $L(E)$ can therefore be described by calculating
\begin{align}
  &\frac{\partial^2}{\partial E^2} L(E) \notag\\
  &~= \frac{\partial^2}{\partial E^2} \Big\{ \smashdiagram{L2} + \smashdiagram{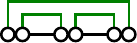} + O(G^4) \Big\} \\
  &~= \smashdiagram{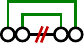} + 2 \smashdiagram{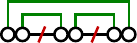} + O(G^4) \,.
  \label{eq:L-diagrams-1}
\end{align}
In these diagrams the slash indicates an energy derivative and the factor $2$ comes from the $2!$ possible orders of taking the two derivatives.
All resolvent lines here and in the following represent effective resolvents [\Eq{eq:effective-resolvent}].
It should be noted that here we do not explicitly write the symmetry factors in the diagrams in contrast to Ref.~\cite{Reininghaus14}.
The vertices in \Eq{eq:L-diagrams-1} are effective vertices, such that we can exclude diagrams like $\smashdiagram{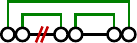}$.

The energy derivatives in \Eq{eq:L-diagrams-1} split the diagrams in such a way that each vertex or block of vertices between two energy derivatives can be replaced by an effective vertex. Due to the energy derivatives this does not lead to double counting of diagrams.
By using effective instead of bare vertices \Eq{eq:L-diagrams-1} includes a large class of diagrams in bare perturbation theory that contribute to $\partial_E^2 L(E)$.
The same logic applies to other diagrams including energy derivatives that form the starting point of our calculation. In the following all vertices in diagrams therefore represent effective vertices.

A fundamental difference between the resolvent and the vertices in renormalized perturbation theory should be emphasized.
We used the effective resolvents instead of bare ones \emph{before} taking the energy derivative without causing double counting. But the effective vertices can only be introduced \emph{after} taking the $E$-derivatives as a simplified notation for a collection of many diagrams in bare perturbation theory.

Since now $\partial_E^2 L(E)$ depends also on the energy-dependent effective vertices, we also need an RG equation describing the energy dependence of these vertices. For the diagrams defining the effective vertices, a single energy derivative is sufficient to achieve convergence in the limit $D\to\infty$ \cite{Reininghaus14}:
\begin{multline}
  \frac{\partial}{\partial E} \Gfull_{12}^{p_1p_2}(E;\bar\omega_1,\bar\omega_2) =
  \diagram{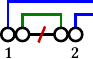} + \diagram{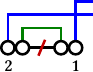} \\
  + \diagram{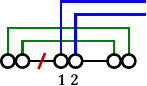} + \diagram{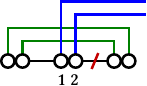} \\
  + \bigg[ \diagram{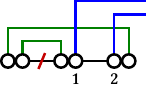} + \diagram{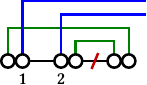} - (1\leftrightarrow2) \bigg] \\
  + O(G^4) \,.
  \label{eq:G-diagrams-1}
\end{multline}

\subsection{Derivatives of contraction lines}

The derivatives in these diagrams can be used to simplify the computation of the diagrams.
The derivative acts on the resolvent in the form $\partial_E R_{12}(E) = \partial_E R(E-\hat{\bar\mu}_{12}+\bar\omega_{12}) = \partial_{\bar\omega_1} R_{12}(E)$. By taking the derivative with respect to the frequencies instead of $E$ we can use integration by parts and take the derivative of the contraction lines and effective vertices instead of derivatives of resolvents. For example, the first diagram in \Eq{eq:G-diagrams-1} can be written as
\begin{equation}
  \diagram{G2slash1} = -\,\diagram{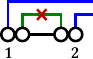} - \diagram{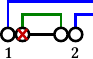} - \diagram{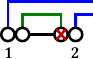}\,.
\end{equation}
Here the cross denotes a frequency derivative of the contraction $\partial_{\bar\omega_3}\gamma^p(\bar\omega_3)$ or of the vertex $\partial_{\bar\omega_3} \Gfull^{p_1p_3}_{13}(E;\bar\omega_1,\bar\omega_3)$.
Since we consider the zero temperature limit, the contraction simplifies to $\partial_\omega \gamma^p(\omega) = -\delta(\omega)$ [\Eq{eq:reservoir-contraction-T0}].

The frequency derivative of the effective vertex can be calculated using the definition of $\Gfull^{pp'}_{11'}(E;\omega,\omega')$ in terms of diagrams with bare vertices.
To this end we consider the second order diagrams for the effective vertex [\Eq{eq:effective_vertex}]. Here one should carefully check by which frequency one differentiates and in which resolvent this frequency enters.
The resulting vertex derivatives,
\begin{align}
  \diagram{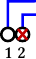} &= \diagram{G2slash2} + O(G^3) \,, \\
  \diagram{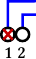} &= \diagram{G2slash1} + O(G^3) \,, \\
  \diagram{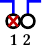} &= \diagram{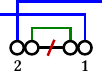} + O(G^3) \,,
\end{align}
are sufficient to compute $\partial_E \Gfull^{pp'}_{11'}(E;\omega,\omega')$ to third order in the vertex:
\begin{multline}
  \diagram{G2slash1} = - \diagram{G2cross1} - \diagram{G3slash3} \\ - \diagram{G3slash4} + O(G^4) \,.
  \label{eq:G_slash}
\end{multline}

Analogously, by using integration by parts twice, one obtains
\begin{align}
  \!\smashdiagram{L2doubleslash}
  &= \smashdiagram{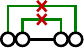} + 2 \smashdiagram{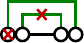} + 2 \smashdiagram{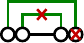}
  \notag\\&\hspace{.5\linewidth} + O(G^4) \\
&= \smashdiagram{L2doublecross} - 2 \smashdiagram{L32slash} + O(G^4) \,.
  \label{eq:L_doubleslash}
\end{align}
Here it is important to remember that all diagrams except for the last one include a prefactor $\frac12$ because two vertices are directly connected by two lines.
Inserting Eqs.~\eqref{eq:G_slash} and \eqref{eq:L_doubleslash} in Eqs.~\eqref{eq:L-diagrams-1} and \eqref{eq:G-diagrams-1} one finds that many terms cancel and we obtain
\begin{align}
  \frac{\partial^2}{\partial E^2} L(E) &= \smashdiagram{L2doublecross} + O(G^4) \,, \label{eq:L-diagrams-2} \\
  \frac{\partial}{\partial E} \Gfull_{12}^{p_1p_2}(E;\bar\omega_1,\bar\omega_2) &=
  - \diagram{G2cross1} - \diagram{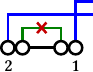} \notag \\
  &\quad - \diagram{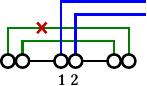} + O(G^4) \,.
  \label{eq:G-diagrams-2}
\end{align}
For the last diagram we have used integration by parts once again, noting that derivatives of the effective vertices in this last diagram would only contribute terms of the order $O(G^4)$.
Here we emphasize that the last diagram still includes a symmetry factor $\frac12$ because two reservoir contraction lines directly connect the same vertices. The frequency derivative (cross) in the contraction line does not change this rule.

\subsection{Frequency dependence of the vertices}
So far, the effective vertex $\Gfull^{pp'}_{11'}(E;\omega,\omega')$ still  depends on two frequencies. In order to derive a numerically solvable  set of differential equations describing the energy dependence of $L(E)$, we need to get rid of this frequency dependence.
Fortunately, at zero temperature most frequencies in the diagrams in Eqs.~\eqref{eq:L-diagrams-2} and \eqref{eq:G-diagrams-2} are set to zero by the contraction $\partial_{\bar\omega} \gamma^p(\bar\omega) = -\delta(\bar\omega)$.
In the diagrammatic language we represent vanishing frequencies in the effective vertices by filled circles.
By including only the vertex at vanishing frequencies in the RG equations we find
\begin{align}
  \frac{\partial^2}{\partial E^2} L(E) &= \smashdiagram{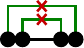} + O(G^4) \,, \label{eq:L-diagrams-3} \\
  \frac{\partial}{\partial E} \Gfull_{12}^{p_1p_2}(E) &=
  - \diagram{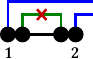} - \diagram{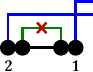} \notag\\
  &\quad - \diagram{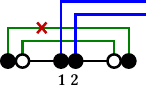} + O(G^4) \,,
  \label{eq:G-RGeq-frequencies}
\end{align}
where we introduced the notation $\Gfull_{12}^{p_1p_2}(E)\coloneqq \Gfull_{12}^{p_1p_2}(E,\bar\omega_1=0,\bar\omega_2=0)$ for vertices at zero frequencies.
Here the last diagram still includes the frequency dependent vertex.
However, the definition of the effective vertex in \Eq{eq:effective_vertex} reminds us that the frequency dependence of the effective vertex is weak and only enters in the next-to-leading order, $\Gfull^{pp'}_{11'}(E;\omega,\omega')=\Gfull^{pp'}_{11'}(E)+O(G^2)$.
Thus we can neglect the frequency dependence of the effective vertices in the last diagram of \Eq{eq:G-RGeq-frequencies} to obtain a closed set of differential equations.

\subsection{Frequency integral}
In the last diagram in \Eq{eq:G-RGeq-frequencies} one integral over an internal frequency remains.
As we have already neglected the frequency dependence of the effective vertices, only the contraction and the resolvents depend on this internal frequency.
To calculate this integral analytically we approximate the frequency dependence of the resolvent $R_{34}(E)$. We use the short-hand notation $\hat{E}_X\coloneqq E-\hat{\bar\mu}_X$ where $X$ denotes a set of indices, \textit{e.g.} $X=34$.
The frequency dependence of the Liouvillian $L(\hat{E}_X+\bar\omega_X)$ can be linearized around $\bar\omega_X=0$ with corrections of order $O(G^2)$ to obtain the approximation
\begin{align}
  R_X(E) &= \frac1{\hat{E}_X + \bar\omega_X + \hat N\Omega - L(\hat{E}_X + \bar\omega_X)} \label{eq:resolvent-frequency-approx-full} \\
  &= \frac1{\bar\omega_X + \chi(\hat{E}_X)} Z(\hat{E}_X) + O(G^2) \,, \label{eq:resolvent-frequency-approx} \\
\chi(\hat{E}_X) &= Z(\hat{E}_X) [\hat{E}_X + \hat N\Omega - L(\hat{E}_X)] \,, \\
  Z(\hat{E}_X) &= \frac1{1-\frac{\partial}{\partial E} L(\hat{E}_X)} \,. \label{eq:Z-definition}
\end{align}
This leading order approximation is sufficient to evaluate the frequency integral to third order in the vertices, consistent with the truncation order used before.

Inserting the approximated resolvent \eqref{eq:resolvent-frequency-approx} in the diagrams of \Eq{eq:G-RGeq-frequencies} yields in the limit of $D\to\infty$:
\begin{widetext}
\begin{multline}
  \frac{\partial}{\partial E} \Gfull^{p_1p_2}_{12}(E)
  = \Gfull^{p_1p_3}_{13}(E) \frac1{\hat{E}_{13} + \hat N \Omega - L(\hat{E}_{13})} \Gfull^{p_4 p_2}_{\bar32}(\hat{E}_{13}) \,-\, (1\leftrightarrow2) \\
  + \frac12 \Gfull^{p_3 p_4}_{34}(E) \int d\omega_4\, p_5\Theta(-p_5\bar\omega_4) \frac1{\bar\omega_4+\chi(\hat{E}_{34})} Z(\hat{E}_{34}) \Gfull^{p_1 p_2}_{12}(\hat{E}_{34}) \frac1{\bar\omega_4+\chi(\hat{E}_{1234})}Z(\hat{E}_{1234}) \Gfull^{p_5p_6}_{\bar4\bar3}(\hat{E}_{1234}) + O(G^4)
  \label{eq:G-integral-1}
\end{multline}
The integral in \Eq{eq:G-integral-1} could be solved exactly by diagonalizing $\chi(\hat{E}_{34})$ and $\chi(\hat{E}_{1234})$ numerically.
But this is computationally expensive and can be avoided by approximating the integral, using that the dependence of the integral on the bias voltage $\hat{\bar\mu}_{12}$ is weak.
As we show in App.~\ref{app:integral}, the second line in \Eq{eq:G-integral-1} equals
\begin{equation}
  -\frac14 \Gfull^{p_3p_4}_{34}(E) \left[ \frac1{\chi(\hat{E}_{34})} Z(\hat{E}_{34}) \Gfull^{p_1p_2}_{12}(\hat{E}_{34}) + Z(\hat{E}_{34}) \Gfull^{p_1p_2}_{12}(\hat{E}_{34}) \frac1{\chi(\hat{E}_{1234})} \right] Z(\hat{E}_{1234}) \Gfull^{p_5p_6}_{\bar4\bar3}(\hat{E}_{1234}) + O\Big(\frac{\Delta G^3}{{\tilde E}^2}\Big) + O(G^4) \,. \label{eq:G-integral-2}
\end{equation}
Here $\Delta\sim\mu_{LR}$ denotes the energy scale of the bias voltage and $\tilde{E}\sim E-L(E)$ is a renormalized energy scale that remains finite
for $E\to0$.
The selection of diagrams and approximations will be discussed further in \autoref{sec:diagram-selection}.

Having solved the integral, we can summarize the RG equations for the effective Liouvillian and the vertex:
\begin{align}
  \frac{\partial^2}{\partial E^2} L(E)
  &= \frac12 \Gfull^{p_1p_2}_{12}(E) R(\hat{E}_{12}) \Gfull^{p_3p_4}_{\bar2 \bar1}(\hat{E}_{12}) + O(G^4) \,, \label{eq:L-RGeq-general-eta} \\
\frac{\partial}{\partial E} \Gfull^{p_1p_2}_{12}(E)
  &= \Gfull^{p_1p_3}_{13}(E) R(\hat{E}_{13}) \Gfull^{p_4p_2}_{\bar32}(\hat{E}_{13}) - (1\leftrightarrow2)
  \notag \\ &~
  - \frac14 \Gfull^{p_3p_4}_{34}(E) \Big[ R(\hat{E}_{34}) \Gfull^{p_1p_2}_{12}(\hat{E}_{34}) Z(\hat{E}_{1234})
+ Z(\hat{E}_{34}) \Gfull^{p_1p_2}_{12}(\hat{E}_{34}) R(\hat{E}_{1234}) \Big] \Gfull^{p_5p_6}_{\bar4\bar3}(\hat{E}_{34})
  \notag \\ &~
  + O(\Delta G^3{\tilde E}^{-2}) + O(G^4) \,.
  \label{eq:G-RGeq-general-eta}
\end{align}
\null
\end{widetext}
Again, summation over all indices that do not appear on the left hand side of the equation is implicit.

\subsection{Summing over $p$ and $\eta$ indices}
The RG equations~\eqref{eq:L-RGeq-general-eta} and \eqref{eq:G-RGeq-general-eta} show that only the vertex averaged over the Keldysh indices is required to compute $L(E)$.
We can therefore define $\Gkavg_{11'}(E)\coloneqq \sum_{pp'}\Gfull^{pp'}_{11'}(E)$ and ignore all Keldysh indices in the RG equations.

The RG equations can be simplified further by removing the index $\eta$ from the multi-index $1$.
When defining the vertex in \Eq{eq:vertex-liouvillian} we have restricted it to the form $\Gfull^{pp'}_{11'}=-\Gfull^{p'p}_{1'1}$.
Furthermore, we consider only particle number conserving vertices that fulfill $\Gkavg_{11'}\propto\delta_{\eta\bar\eta'}$.
These symmetries allow us to include only vertices $\Gkavg_{11'}(E)$ with indices $\eta=-\eta'=+$ in the RG equations. All other index combinations are either redundant or not allowed by particle number conservation. The vertices with $\eta=-\eta'=+$ will be denoted by $\Gnoeta_{11'}$ where the multi-indices $1,1'$ do not include $\eta$ anymore.
Correspondingly, when $\eta$ is removed from the index $1$, we define $\hat{\bar\mu}_{12}=\hat\mu_1-\hat\mu_2$.
To illustrate how this can be used in the RG equations we schematically write the right hand side of the RG equation for $L(E)$ and ignore all indices except for $\eta$: $\sum_{\eta\eta'}\Gkavg_{\eta\eta'}R\Gkavg_{\bar\eta'\bar\eta}$. Using that $\Gkavg_{\eta\eta'}\propto\delta_{\eta\bar\eta'}$ and that $\Gkavg_{-+}=-\Gkavg_{+-}$ we can write this as $2\Gkavg_{+-}R\Gkavg_{+-}$.
With these considerations we obtain the RG equations
\begin{align}
  \frac{\partial^2}{\partial E^2} L(E)
  &= \Gnoeta_{12}(E) R(\hat{E}_{12}) \Gnoeta_{21}(\hat{E}_{12}) + O(G^4) \,, \label{eq:L-RGeq-general} \\
  \frac{\partial}{\partial E} \Gnoeta_{12}(E)
  &= \Gnoeta_{13}(E) R(\hat{E}_{13}) \Gnoeta_{32}(\hat{E}_{13})
  \notag \\ &\quad
  - \Gnoeta_{32}(E) R(\hat{E}_{32}) \Gnoeta_{13}(\hat{E}_{32})
  \notag \\ &\quad
  - \frac12 \Gnoeta_{34}(E) \Big[ R(\hat{E}_{34}) \Gnoeta_{12}(\hat{E}_{34}) Z(\hat{E}_{1234})
  \notag \\ &\quad\hphantom{-}
  + Z(\hat{E}_{34}) \Gnoeta_{12}(\hat{E}_{34}) R(\hat{E}_{1234}) \Big] \Gnoeta_{43}(\hat{E}_{34})
  \notag \\ &\quad
  + O(\Delta G^3{\tilde E}^{-2}) + O(G^4) \,.
  \label{eq:G-RGeq-general}
\end{align}

\subsection{Observables}
\label{sec:RGeq-observables}
The effective Liouvillian describes the evolution of the system without the reservoirs, which is only the central spin. To compute observables like the current, we need to include these observables in the RG equations and in the diagrammatic language.
The diagrammatic representation of observables can be derived by using a close analogy to the diagrammatic representation of the time derivative of a state.
Writing $\rho^\mathrm{tot}(t)$ for the density matrix of the full system (including reservoirs) at time $t$, we can express the derivative of the density matrix $\rho(t)=\tr_R \rho^\mathrm{tot}(t)$ as [cf.~\Eq{eq:pi-time-domain}]
\begin{align}
  \frac{d}{dt} \rho(t)
&= \tr_R \left\{-iL_V\rho^\mathrm{tot}(t)\right\} \label{eq:L-time-domain-0} \\
  &= \int_{t_0}^t dt'\, L(t,t') \rho(t')\,.
  \label{eq:L-time-domain}
\end{align}
\Eq{eq:L-time-domain} follows from the time domain equivalent to \Eq{eq:L-floquet-space}.
The evolution of an observable represented by an operator $A$ can be calculated in a similar way. We define a superoperator $\mathcal{A}=\frac12\{A,\bullet\}$ to write the expectation value of $A$ at time $t$ in the form
\begin{align}
  \langle A\rangle(t)
  &= \tr \tr_R \left\{\mathcal{A}\rho^\mathrm{tot}(t)\right\} \label{eq:observable-kernel-time-domain-0} \\
  &= \tr \int_{t_0}^t dt'\, \Sigma_A(t,t') \rho(t')
  \label{eq:observable-kernel-time-domain}
\end{align}
where the last step defines the kernel $\Sigma_A$.
The close analogy of Eqs.~\eqref{eq:L-time-domain-0} and \eqref{eq:observable-kernel-time-domain-0} suggests that $L$ and $\Sigma_A$ can be computed in a similar way.
Indeed, when $A$ is bilinear in the fermion creation and annihilation operators, the expansions of $L$ and $\Sigma_A$ in orders of $L_V$ can be compared term by term. For each term in the expansion of $L$ the corresponding term in the expansion of $\Sigma_A$ can be obtained by replacing the left-most occurrence of $L_V$ by $\mathcal{A}$.
For the diagrammatic language this implies that by defining a vertex for $\mathcal{A}$ similar to the coupling vertex for $L_V$ we can use the same diagrams for $L$ and $\Sigma_A$ with the only difference that the left-most vertex in the diagrams for $\Sigma_A$ must be the vertex representing $\mathcal{A}$ instead of the coupling vertex.
We will use this analogy in the diagrammatic language to compute the current. A more detailed derivation of the diagrammatic language for observables can be found in Ref.~\cite{Schoeller09}.

\subsubsection{Current}
The current in lead $\gamma$ is described by the operator $I^\gamma=-i[H,N_\gamma]$ which is bilinear in the reservoir creation and annihilation operators. Here $N_\gamma$ is the particle number operator in reservoir $\gamma$.
In order to include this observable in the diagrammatic language we define a current vertex $\Ifull^{\gamma\,pp'}_{11'}$ analogous to \Eq{eq:vertex-liouvillian} by writing
\begin{equation}
  \mathcal{A} \coloneqq i\frac12\left\{ I^\gamma, \bullet \right\} = \frac12 p' \Ifull^{\gamma\,pp'}_{11'} :J^p_1 J^{p'}_{1'}:
  \label{eq:current-vertex}\,.
\end{equation}
Like the coupling vertex, the current vertex should be defined such that $\Ifull^{\gamma\,p'p}_{1'1}(E)=-\Ifull^{\gamma\,pp'}_{11'}(E)$.
The corresponding current kernel $\Sigma_A\eqqcolon\Sigma_\gamma$ defined in \Eq{eq:observable-kernel-time-domain} can be calculated by using the diagrams for $L$ and replacing in each diagram the left-most coupling vertex by a current vertex.
Since we are eventually only interested in $\tr\Sigma_\gamma(E)$, we can neglect all terms which do not contribute to the trace in the current kernel $\Sigma_\gamma(E)$ and in the vertex $\Ifull^{\gamma\,pp'}_{11'}$.

In analogy to the coupling vertex we define an effective current vertex $\Inoeta_{12}$ averaged over Keldysh indices and with fixed $\eta_1=-\eta_2=+$ to write the RG equation for $\Sigma_\gamma$ in the same compact form as \Eq{eq:L-RGeq-general} for $L(E)$:
\begin{equation}
  \frac{\partial^2}{\partial E^2} \Sigma_\gamma(E)
  = \Inoeta^\gamma_{12}(E) R(\hat{E}_{12}) \Gnoeta_{21}(\hat{E}_{12}) + O(G^4) \,.
\end{equation}
Analogously, the RG equation for the current vertex $\Inoeta^\gamma_{12}$ is obtained from the RG equation for the coupling vertex $\Gnoeta_{12}(E)$ [\Eq{eq:G-RGeq-general}] by replacing the leftmost coupling vertex in each diagram by a current vertex.

\subsubsection{Differential conductance}

Besides the current we also include the differential conductance $dI^\gamma/d\vdc$ in the RG equations. Here $\vdc$ is the time-averaged bias voltage $\mu_L-\mu_R$. By using independent RG equations for the current and the differential conductance we obtain a consistency check of our results and can directly compare the FRTRG data to previous RTRG calculations.

To directly calculate the differential conductance we consider the variation of the Liouvillian and the current kernel for infinitesimal changes of $\hat{\bar\mu}_X$ by a scalar, $\delta\hat{\bar\mu}_{LR}=\delta\vdc\,\id$, representing a time-constant variation of the bias voltage.
In the diagrams with bare vertices the variation of $\hat{\bar\mu}$ enters only through the variation of the resolvent,
\begin{equation}
  \delta R_X(E) = \delta\hat{\bar\mu}_X \frac\partial{\partial E} R_X(E) + R_X(E) \delta L(\hat{E}_X+\bar\omega_X) R_X(E) \,.
  \label{eq:variation-resolvent}
\end{equation}
The variation $\delta L(E)=L(E)|_{\hat{\bar\mu}+\delta\hat{\bar\mu}}-L(E)|_{\hat{\bar\mu}}$ of the Liouvillian will be denoted by $\includegraphics{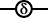}$.
To construct the diagrams for $\partial_E \delta L(E)$ we start from the diagrams for $L(E)$ with bare vertices and apply the variation of $\hat{\bar\mu}$ which acts on a single resolvent as described in \Eq{eq:variation-resolvent}. We then take the $E$ derivative and collect bare vertices in effective vertices to obtain
\begin{multline}
  \frac{\partial}{\partial E} \delta L(E) = \delta\hat{\bar\mu}_{12} \diagram{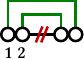} \\
  + \smashdiagram{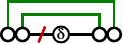} + \smashdiagram{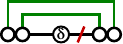} \\ + (\delta\hat{\bar\mu}_{12}+\delta\hat{\bar\mu}_{13}) \diagram{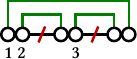} + O(G^4) \,.
\end{multline}
Here we neglected all diagrams of order $G^3 \delta L$ for the following reason. The leading order shows that $\partial_E \delta L(E)=O(G^2)$ just like $\partial_E G(E)=O(G^2)$ [\Eq{eq:G-RGeq-general}]. We will see later [\autoref{sec:initial-conditions-summary}] that also the initial conditions for the variation of $L$ and of $\Sigma_\gamma$ are of order $O(G^2)$. Thus we can conclude that $\delta L=O(G)$ and all diagrams containing $\delta L$ and more than two vertices are negligible.

Analogous to the derivations of Eqs.~\eqref{eq:L-diagrams-2} and \eqref{eq:L-diagrams-3} we find that after integration by parts some terms cancel and the frequency dependence of the vertex can be neglected:
\begin{equation}
  \frac{\partial}{\partial E} \delta L(E)
  = \delta\hat{\bar\mu}_{12} \diagram{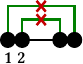} - \smashdiagram{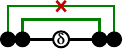} + O(G^4) \\
\end{equation}
The evaluation of the second diagram involves again an integral of the form solved in App.~\ref{app:integral}, which we have already encountered in the RG equations for the coupling vertex [\Eq{eq:G-integral-1}]. The resulting RG equation for $\delta L(E)$ reads
\begin{align}
  \frac{\partial}{\partial E} \delta L(E)
  &=\frac12 \delta\hat{\bar\mu}_{12} \Gkavg_{12}(E) R(\hat{E}_{12}) \Gkavg_{\bar2\bar1}(\hat{E}_{12})
  \notag \\ &\quad
  - \frac14 \Gkavg_{12}(E) \Big[ R(\hat{E}_{12}) \delta L(\hat{E}_{12}) Z(\hat{E}_{12})
  \notag \\ &\quad
  + Z(\hat{E}_{12}) \delta L(\hat{E}_{12}) R(\hat{E}_{12}) \Big] \Gkavg_{\bar2\bar1}(\hat{E}_{12})
  \notag \\ &\quad
  + O(\Delta G^3{\tilde E}^{-2}) + O(G^4)
  \\
  &= \delta\hat{\bar\mu}_{12} \Gnoeta_{12}(E) R(\hat{E}_{12}) \Gnoeta_{21}(\hat{E}_{12})
  \notag \\ &\quad
  - \frac12 \Gnoeta_{12}(E) \Big[ R(\hat{E}_{12}) \delta L(\hat{E}_{12}) Z(\hat{E}_{12})
  \notag \\ &\quad
  + Z(\hat{E}_{12}) \delta L(\hat{E}_{12}) R(\hat{E}_{12}) \Big] \Gnoeta_{21}(\hat{E}_{12})
  \notag \\ &\quad
  + O(\Delta G^3{\tilde E}^{-2}) + O(G^4) \,.
  \label{eq:dL-RGeq-general}
\end{align}
The RG equation for $\delta\Sigma_\gamma$ is obtained from \Eq{eq:dL-RGeq-general} by replacing the leftmost vertex of each diagram by the current vertex.

\section{RG~equations for the Kondo~model}
\label{sec:kondo-RG}

Having derived the RG equations on a general level, we can now consider the Kondo model to provide explicit expressions for the bare vertices and parametrizations of the abstract superoperators appearing in the general RG equations.
Again we closely follow Ref.~\cite{Reininghaus14}.

\subsection{Kondo model in Liouville space}

We begin by rewriting the Kondo model in the language used in the general RG equations.
The coupling Liouvillian $L_V = [V,\bullet]$ is determined by the coupling Hamiltonian
\begin{equation}
  V = \frac12 \iint d\omega d\omega' \sum_{\alpha\alpha'\sigma\sigma'} J_{\alpha\alpha'}^{(0)} \mathbf{S}\cdot\boldsymbol{\sigma}_{\sigma\sigma'} c_{\alpha\sigma\omega}^\dag c_{\alpha'\sigma'\omega'} \,.
\end{equation}
Recalling the definition of the fermion superoperators $J^+_1=c_1\bullet$ and $J^-_1=\bullet c_1$, we can rewrite the coupling Liouvillian in the form
\begin{align}
  L_V &= \frac14 J_{\alpha\alpha'}^{(0)} \eta \delta_{\eta\bar\eta'} \boldsymbol{\sigma}_{\sigma\sigma'} \cdot \left( \mathbf{S} :c_1c_{1'}: \bullet + \bullet \mathbf{S} :c_{1'}c_1: \right) \\
  &= \frac12 p' \Gfull^{(0)pp'}_{11'} :J^p_1 J^{p'}_{1'}:\,,
\end{align}
as required for the diagrammatic language [\Eq{eq:vertex-liouvillian}].
The bare vertex is defined such that $\Gfull^{(0)p'p}_{1'1}=-\Gfull^{(0)pp'}_{11'}$:
\begin{equation}\Gfull^{(0)pp'}_{11'} = \delta_{pp'}
  \begin{cases}
    g_{11'} \bullet & \text{ for } p=+, \\
    - \bullet g_{11'} & \text{ for } p=-, \\
  \end{cases}
  \label{eq:bare-vertex}
\end{equation}
with the operator
\begin{equation}
  g_{11'} = \frac12 \delta_{\eta\bar\eta'}
  \begin{cases}
    J_{\alpha\alpha'}^{(0)} \mathbf{S}\cdot\boldsymbol{\sigma}_{\sigma\sigma'} & \text{ for } \eta=+, \\
    - J_{\alpha'\alpha}^{(0)} \mathbf{S}\cdot\boldsymbol{\sigma}_{\sigma'\sigma} & \text{ for } \eta=-.
  \end{cases}
  \label{eq:small-g}
\end{equation}

The current in lead $\gamma$ can be calculated using $I^\gamma=-d\langle N_\gamma\rangle/dt=-i\langle [V,N_\gamma]\rangle$,
\begin{equation}
  [V,N_\gamma] = -\frac12 J_{\alpha\alpha'}^{(0)} (\delta_{\alpha\gamma}-\delta_{\alpha'\gamma}) \mathbf{S}\cdot\boldsymbol\sigma_{\sigma\sigma'} c_{\alpha\sigma\omega}^\dag c_{\alpha'\sigma'\omega'} \,.
\end{equation}
In Liouville space this observable can be expressed by a current vertex following \Eq{eq:current-vertex} via
\begin{align}
  \frac12 \big\{[V,N_\gamma],\bullet\big\}
  &= \frac12 p'\Ifull_{11'}^{\gamma(0)pp'} :J_1^p J_{1'}^{p'}: \,, \\
  \Ifull_{11'}^{\gamma(0)pp'}
  &= -\frac12 p' \big(\eta\delta_{\alpha\gamma} + \eta'\delta_{\alpha'\gamma}\big) \Gfull_{11'}^{(0)pp'} \,.
  \label{eq:bare-current-vertex}
\end{align}

These bare quantities will be helpful for computing the initial conditions of the RG flow, which will be based on bare perturbation theory. But for the RG equations we additionally need a parametrization of the effective quantities, which we will derive in the following.

\subsection{Parametrization of superoperators}

In order to use the RG equations in practice we need to parametrize the superoperators by scalars in Liouville space, which, however, remain matrices in Floquet space.
In the isotropic Kondo model without magnetic field the evolution of the density matrix $\rho$ describing the state of spin $\mathbf{S}$ must preserve the rotational symmetry, the normalization $\tr\rho=1$, and the hermiticity of the state.
This restricts the effective Liouvillian $L$ to the form
\begin{equation}
  L(E) = -i\Gamma(E) L^a \label{eq:L-parametrization}
\end{equation}
with the scalar spin relaxation rate $\Gamma$ and the constant superoperator $L^a$ defined by
\begin{equation}
  L^a\rho = \rho - \tfrac12 \id \tr \rho \,. \label{eq:La}
\end{equation}
In the diagrammatic language and in the RG equations $L(E)$ acts only on traceless operators, because the coupling vertex is traceless: $\tr \Gkavg_{11'}=0$.
Thus we can simply replace the superoperator $L(E)$ by the scalar $-i\Gamma(E)$. With this replacement also the resolvent $R(E)$ and $Z(E)$ become scalars in superoperator space.

Also the current kernel $\Sigma_\gamma$ can be reduced to a scalar. Since we consider the isotropic Kondo model, the state of the central spin will always be unpolarized after tracing out the reservoirs. As $\Sigma_\gamma$---by construction of the diagrammatic expansion---only acts on the state when the reservoirs have just been reset, the only operator on which $\Sigma_\gamma$ will ever act is the unpolarized spin state $\frac12 \id$. Furthermore, for calculating the current we only need the trace $\tr\Sigma_\gamma\frac12\id$ of the current kernel, such that only a single relevant matrix element of the superoperator $\Sigma_\gamma$ remains.
We parametrize the current kernel as
\begin{align}
  \Sigma_\gamma(E) &= i\Gamma^\gamma(E) L^b \,, \label{eq:current-kernel-parametrization} \\
  L^b &= \frac12\id\tr=1-L^a \,. \label{eq:Lb}
\end{align}

A similar simplification can be done for the current vertex $\Inoeta^\gamma_{11'}$.
To compute the current we only need $\tr \Inoeta_{11'}^\gamma$, and $\Inoeta_{11'}^\gamma$ only acts on traceless operators in the diagrammatic language and in the RG equations. As we formally derive in App.~\ref{app:symmetries}, the rotational symmetry in spin space restricts the parametrization further to the form
\begin{align}
  \Inoeta_{11'}^\gamma &= -\frac14 I_{11'}^\gamma \hat{L}^1_{\sigma\sigma'} \,,
  \label{eq:I-parametrization} \\
  \hat{L}^1_{\sigma\sigma'} &= \frac12 \sigma^i_{\sigma\sigma'} \id \tr(\sigma_i \bullet) \,,
  \label{eq:L1-def}
\end{align}
where $I_{11'}^\gamma$ is the scalar current vertex.

A more involved parametrization is required for the coupling vertex $\Gnoeta_{12}(E)$.
It must preserve the rotational symmetry and hermiticity, and ensure that the resulting diagrams are traceless, $\tr \Gnoeta_{11'}(E)=0$.
In App.~\ref{app:symmetries} we prove using these restrictions that the vertex can be parametrized by three scalars in Liouville space:
\begin{align}
  \Gnoeta_{11'}(E) &= \sum_{\chi=a,2,3} G^\chi_{11'}(E) \hat{L}^\chi_{\sigma\sigma'} \,, \label{eq:G-parametrization} \\
  \hat{L}^a_{\sigma\sigma'} &= \delta_{\sigma\sigma'} L^a \,, \\
  \hat{L}^2_{\sigma\sigma'} &= -\frac14 \sum_{i=1}^3 \sigma^i_{\sigma\sigma'} [\sigma_i, \bullet] \,, \\
  \hat{L}^3_{\sigma\sigma'} &= \frac12 \sum_{i=1}^3 \sigma^i_{\sigma\sigma'} \sigma_i\tr \,.
\end{align}
Writing the bare vertex in this form,
\begin{equation}
  \sum_{pp'} \Gfull^{(0)pp'}_{11'}
= \frac12 J_{\alpha\alpha'}^{(0)} [\mathbf{S}\cdot\boldsymbol\sigma_{\sigma\sigma'}, \bullet]
  = -J_{\alpha\alpha'}^{(0)} \hat{L}^2_{\sigma\sigma'} \,,
  \label{eq:bare-vertex-leading-L2}
\end{equation}
shows that only $G^2$ appears in the bare vertex and is thereby the leading order contribution to the vertex. In our RG method that is based on an expansion in orders of the coupling we should therefore use $G^2$ as the reference scale that we will denote by $G^2=O(J)$.
While $G^3=O(J^2)$ is essential for calculating observables, the contribution of $G^a$ to the RG equations is of a negligible higher order as we will see in \autoref{sec:rg-eq-scalars}.

The superoperator algebra required when working with the RG equations can be summarized in Dirac notation, where for any operator $a$ (\textit{e.g.} a density matrix) $\Bra{a}\coloneqq \tr(a^\dag \bullet)$ represents a left vector and an operator $\Ket{a}\coloneqq a$ is interpreted as a right vector:
\begin{align}
  \hat{L}^a_{\sigma\sigma'} &= \delta_{\sigma\sigma'} L^a = \delta_{\sigma\sigma'} \frac12 \sum_{i=1}^3 \Ket{\sigma_i}\Bra{\sigma_i} \,, \label{eq:La-hat} \\
  \hat{L}^b_{\sigma\sigma'} &= \delta_{\sigma\sigma'} L^b = \frac12 \delta_{\sigma\sigma'} \Ket\id \Bra\id \,, \\
  \hat{L}^1_{\sigma\sigma'} &= \frac12 \sum_{i=1}^3 \sigma^i_{\sigma\sigma'} \Ket\id \Bra{\sigma_i} \,, \\
  \hat{L}^2_{\sigma\sigma'} &= \frac{i}{4} \sum_{i,j,k=1}^3 \varepsilon_{ijk} \sigma^k_{\sigma\sigma'} \Ket{\sigma_i}\Bra{\sigma_j} \,, \\
  \hat{L}^3_{\sigma\sigma'} &= \frac12 \sum_{i=1}^3 \sigma^i_{\sigma\sigma'} \Ket{\sigma_i}\Bra{\id} \label{eq:L3-hat} \,.
\end{align}

For the RG equations we will need the products $(\hat{L}^i \hat{L}^j)_{\sigma_1\sigma_2} = \sum_{\sigma_3} \hat{L}^i_{\sigma_1\sigma_3} \hat{L}^j_{\sigma_3\sigma_2}$ and $(\hat{L}^{i\,\top} \hat{L}^{j\,\top})^\top_{\sigma_1\sigma_2} = \sum_{\sigma_3} \hat{L}^i_{\sigma_3\sigma_2} \hat{L}^j_{\sigma_1\sigma_3}$ where the transpose is only taken in reservoir spin space.
The multiplication table for these products is given in \autoref{tab:multiplication-table}.
It will be helpful to remember that $\hat{L}^b$ and $\hat{L}^3$ vanish when acting on traceless operators while $\hat{L}^a$ leaves all traceless operators unchanged.
This implies that in the RG equations $G^3$ can only stand on the right of all vertices.

\begin{table}[t]
  \setlength\tabcolsep{0.9em}\begin{tabular}{c|ccccc}
    \toprule
    & $\hat{L}^a$ & $\hat{L}^b$ & $\hat{L}^1$ & $\hat{L}^2$ & $\hat{L}^3$ \\
    \midrule
    $\hat{L}^a$ & $\hat{L}^a$ & $0$ & $0$ & $\hat{L}^2$ & $\hat{L}^3$ \\
    $\hat{L}^b$ & $0$ & $\hat{L}^b$ & $\hat{L}^1$ & $0$ & $0$ \\
    $\hat{L}^1$ & $\hat{L}^1$ & $0$ & $0$ & $\pm \hat{L}^1$ & $3\hat{L}^b$ \\
    $\hat{L}^2$ & $\hat{L}^2$ & $0$ & $0$ & $\frac12(\hat{L}^a \pm \hat{L}^2)$ & $\pm \hat{L}^3$ \\
    $\hat{L}^3$ & $0$ & $\hat{L}^3$ & $\hat{L}^a \pm 2\hat{L}^2$ & $0$ & $0$ \\
    \bottomrule
  \end{tabular}
  \caption{\label{tab:multiplication-table}Products $\hat{L}^i \hat{L}^j$ (upper sign) and $(\hat{L}^{i\,\top} \hat{L}^{j\,\top})^\top$ (lower sign), where the transpose only affects the spin indices, $\hat{L}^{i\,\top}_{\sigma\sigma'}=\hat{L}^i_{\sigma'\sigma}$.
    See Tables~I and II in Ref.~\cite{Reininghaus14}.
  }
\end{table}

\subsection{Diagram selection}
\label{sec:diagram-selection}
In \autoref{sec:general-RG} the general RG equations have been derived up to next-to-leading order in the coupling vertex, which defines the scale $J\sim G^2_{12}$ that serves as our expansion parameter.
The RG equations show that $\partial_E J \sim J^2/\tilde{E}$ and thus $\partial_E J^n = J^{n+1}/\tilde{E}$ with $\tilde{E}=E+i\Gamma(E)$ for the Kondo model.
When setting up the RG equations for the Kondo model we will select terms up to next-to-leading order in $J$ for the previously defined parametrizations.
These include terms up to order $O(J^3/\tilde{E})$ for $G^2$ and $O(J^4/\tilde{E})$ for $G^3$.

Beyond the expansion in orders of $J$ we use one further approximation.
When calculating the integral in the next-to-leading order contribution to $\partial_E G(E)$ in \Eq{eq:G-integral-1} we have already made the assumption that $\Delta/\tilde{E}$ is small. Again $\Delta$ denotes the energy scale of the bias voltage.
In general we make the approximation that for an RG equation with leading order $J^n/\tilde{E}$ we neglect not only $O(J^{n+2}/\tilde{E})$, but also $O(\Delta  J^{n+1}/\tilde{E}^2)$.
This simplifies the RG equations significantly not only in the frequency integral in \Eq{eq:G-integral-1}, but also in the parametrization of the RG equations as we will see in \autoref{sec:rg-eq-scalars}.

To understand why we can make this approximation we distinguish the regimes of weak and strong coupling.
When the bias voltage is small, $\Delta<\tkv$, we are in the strong coupling regime and $\Delta/\tilde{E}$ remains small throughout the RG flow, thereby justifying the approximation.
In the opposite regime of weak coupling the bias voltage is large, $\Delta>\tkv$, the Kondo resonance is weakened, and the effective coupling $J$ is smaller than in the strong coupling regime.
When the bias voltage is the dominant energy scale, it will lead to an increased spin relaxation rate $\Gamma(E=0)$ such that $\Delta/\tilde{E}$ will not increase drastically for small $E$.
Thus, by neglecting $O(\Delta J^{n+1}/\tilde{E}^2)$ we do lose accuracy in the next-to-leading order in $J$ when reaching $|E|<\Delta$ in the RG flow. But since in this regime $J$ is small and only the next-to-leading order in $J$ is affected, this approximation remains justified also for weak coupling.
In App.~\ref{app:g-self-consistency} (\autoref{fig:order-comparison}) we check the validity of this argument by comparing our results to an FRTRG calculation in which we keep only the leading order in $J$.

\subsection{RG equations for scalars}
\label{sec:rg-eq-scalars}
Based on the described strategy for diagram selection we derive the RG equations for the parametrizations of superoperators from the general RG equations.
The RG equation for $\Gamma(E)$ is derived by starting from \Eq{eq:L-RGeq-general} and inserting the parametrization:
\begin{align}
  \frac{\partial^2}{\partial E^2} L(E)
  &= \Gnoeta_{12}(E) R(\hat{E}_{12}) \Gnoeta_{21}(\hat{E}_{12}) \\
  &= \tr_\sigma(\hat{L}^2 \hat{L}^2) G^2_{12}(E) R(\hat{E}_{12}) G^2_{21}(\hat{E}_{12}) \,.
\end{align}
Here $\tr_\sigma$ denotes the trace only over the reservoir spin space, i.e., $\tr_\sigma \hat{L}^i = \sum_\sigma \hat{L}^i_{\sigma\sigma}$, and the spin index $\sigma$ in the multi-index $1$ is absorbed in the algebra of $\hat{L}^i$.
To understand why only one term contributes we look at the algebra of the other terms. For the contribution $(G^2,G^3)$ of the two vertices we have $\tr_\sigma \hat{L}^2\hat{L}^3 = \tr_\sigma \hat{L}^3 = 0$. The $G^3$ contribution of the first vertex cannot contribute because $\hat{L}^3$ vanishes when acting on traceless states, the combination $(G^aG^a)$ is of order $O(J^4)$, and the remaining contributions vanish when tracing over the spin index, $\tr_\sigma \hat{L}^2 \hat{L}^a = \tr_\sigma \hat{L}^a \hat{L}^2 = \tr_\sigma \hat{L}^2 = 0$.
This leads to the RG equation for $\Gamma$,
\begin{equation}
  -i\frac{\partial^2}{\partial E^2} \Gamma(E)
  = G^2_{12}(E) \Ra(\hat{E}_{12}) G^2_{21}(\hat{E}_{12}) \,.
\end{equation}

As we have seen in \autoref{sec:RGeq-observables}, the RG equation for the current kernel $\Sigma_\gamma$ can be obtained from the RG equation~\eqref{eq:L-RGeq-general} for $L$ by just replacing the leftmost coupling vertex by the current vertex.
Since $\Sigma_\gamma\propto L^b\propto\tr$ involves the trace of the input state, only the contribution of $\hat{L}^3\propto\id\tr$ can contribute in the rightmost vertex of the diagram.
Inserting the parametrization we find
\begin{equation}
  i\frac{\partial^2}{\partial E^2} \Gamma^\gamma(E) L^b
  = -\frac14 \tr_\sigma(\hat{L}^1\hat{L}^3) I^\gamma_{12}(E) \Ra(\hat{E}_{12}) G^3_{21}(\hat{E}_{12}) \,,
\end{equation}
which simplifies to
\begin{equation}
  \frac{\partial^2}{\partial E^2} \Gamma^\gamma(E)
  = i\frac32 I^\gamma_{12}(E) \Ra(\hat{E}_{12}) G^3_{21}(\hat{E}_{12}) \,.
\end{equation}

Next we derive the RG equations for the vertex parametrization and start by showing that the contribution of $G^a$ to the RG equations can be neglected.
The leading order contribution to $\partial_E G^a$ in the RG equation~\eqref{eq:G-RGeq-general} for $\Gnoeta_{12}$ only includes $G^2$:
\begin{multline}
  \frac{\partial}{\partial E} G^a_{12}(E) = \frac12 G^2_{13}(E) R(\hat{E}_{13}) G^2_{32}(\hat{E}_{13}) \\ - \frac12 G^2_{32}(E) R(\hat{E}_{32}) G^2_{13}(\hat{E}_{32}) \,.
  \label{eq:G-RGeq-general-ordered}
\end{multline}
For the second term we used $\hat{L}^a_{\sigma_3\sigma_2}\hat{L}^a_{\sigma_1\sigma_3}=(\hat{L}^{a\,\top}\hat{L}^{a\,\top})^\top_{\sigma_1\sigma_2}$ to apply the multiplication table in \autoref{tab:multiplication-table}.
If there is no bias voltage, $\hat\mu_L=\hat\mu_R$, the two terms cancel, which lets us conclude that $\partial_E G^a_{12} = O(\Delta J^2/\tilde{E}^2)$ where $\Delta$ is the energy scale of the bias voltage.
Knowing that $\partial_E G^2(E)=O(J^2/\tilde{E})$, we can conclude that $G^a_{12}=O(\Delta J^2/\tilde{E})$.
This implies that the contribution of $G^a$ to the RG equation for $\partial_E G^2$ is of order $O(\Delta J^3/\tilde{E}^2)$, the contribution to $\partial_E G^3$ is of order $O(\Delta J^4/\tilde{E}^2)$, and also in the observables one finds that $G^a$ only leads to corrections beyond next-to-leading order. Therefore we can completely neglect $G^a$.

We continue with the RG equations for $\partial_E G^2$ and $\partial_E G^3$.
Since the contribution of $G^a$ is neglected, the vertices only contribute the $G^2$ and $G^3$ components. Of these, $G^3$ can only stand on the right of all other vertices since $\hat{L}^3$ vanishes when acting on traceless operators.
Thus the rightmost vertex in \Eq{eq:G-RGeq-general} contributes $G^2$ in the RG equation for $\partial_E G^2$ and $G^3$ for $\partial_E G^3$, while all other vertices only contribute the $G^2$ part.
Knowing which terms contribute, we now only need to compute the products in the spin-superoperator algebra, for which we need the following two terms which are not contained in \autoref{tab:multiplication-table}, but derived in App.~\ref{app:superoperator-algebra}:
\begin{align}
  \sum_{\sigma_3\sigma_4} \hat{L}^2_{\sigma_3\sigma_4} \hat{L}^2_{\sigma_1\sigma_2} \hat{L}^2_{\sigma_4\sigma_3}
  &= \frac12 \hat{L}^2_{\sigma_1\sigma_2} \,,
  \label{eq:superoperator-algebra-1} \\
  \sum_{\sigma_3\sigma_4} \hat{L}^2_{\sigma_3\sigma_4} \hat{L}^2_{\sigma_1\sigma_2} \hat{L}^3_{\sigma_4\sigma_3}
  &= - \hat{L}^3_{\sigma_1\sigma_2} \,.
  \label{eq:superoperator-algebra-2}
\end{align}
Using these relations, we can write down the RG equation for $\partial_E G^2$ and $\partial_E G^3$:
\begin{align}
\frac{\partial}{\partial E} G^2_{12}(E)
&= \frac12 G_{13}^2(E) \Ra(\hat{E}_{13}) G^2_{32}(\hat{E}_{13})
\notag \\ &
+ \frac12 G^2_{32}(E) \Ra(\hat{E}_{32}) G^2_{13}(\hat{E}_{32})
\notag \\ &\quad
- \frac14 G^2_{34}(E) \big[ \Ra(\hat{E}_{34}) G^2_{12}(\hat{E}_{34}) Z(\hat{E}_{1234})
\notag \\ &\quad
+ Z(\hat{E}_{34}) G^2_{12}(\hat{E}_{34}) \Ra(\hat{E}_{1234}) \big] G^2_{43}(\hat{E}_{1234}) \,,
\label{eq:G2-RGeq}
\\
\frac{\partial}{\partial E} G^3_{12}(E)
&= G_{13}^2(E) \Ra(\hat{E}_{13}) G^3_{32}(\hat{E}_{13})
\notag \\ &\quad
+ G^2_{32}(E) \Ra(\hat{E}_{32}) G^3_{13}(\hat{E}_{32})
\notag \\ &\quad
+ \frac12 G^2_{34}(E) \big[ \Ra(\hat{E}_{34}) G^2_{12}(\hat{E}_{34}) Z(\hat{E}_{1234})
\notag \\ &\quad
+ Z(\hat{E}_{34}) G^2_{12}(\hat{E}_{34}) \Ra(\hat{E}_{1234}) \big] G^3_{43}(\hat{E}_{1234}) \,.
\label{eq:G3-RGeq}
\end{align}

The RG equation for the current vertex can be derived in close analogy to \Eq{eq:G3-RGeq}.
The general RG equation for the current vertex is obtained from the RG equation for the coupling vertex [\Eq{eq:G-RGeq-general}] by replacing the leftmost vertex by the current vertex. Inserting the parametrization and using
\begin{equation}
  \sum_{\sigma_3\sigma_4} \hat{L}^1_{\sigma_3\sigma_4} \hat{L}^2_{\sigma_1\sigma_2} \hat{L}^2_{\sigma_4\sigma_3}
  = - \hat{L}^1_{\sigma_1\sigma_2}
  \label{eq:superoperator-algebra-3}
\end{equation}
we obtain the RG equation for $\partial_E I^\gamma$:
\begin{align}
  \frac{\partial}{\partial E} I^\gamma_{12}(E)
  &= I^\gamma_{13}(E) \Ra(\hat{E}_{13}) G^2_{32}(\hat{E}_{13})
  \notag \\ &\quad
  + I^\gamma_{32}(E) \Ra(\hat{E}_{32}) G^2_{13}(\hat{E}_{32})
  \notag \\ &\quad
  + \frac12 I^\gamma_{34}(E) \big[ \Ra(\hat{E}_{34}) G^2_{12}(\hat{E}_{34}) Z(\hat{E}_{1234})
  \notag \\ &\quad\hphantom{+}
  + Z(\hat{E}_{34}) G^2_{12}(\hat{E}_{34}) \Ra(\hat{E}_{1234}) \big] G^2_{43}(\hat{E}_{1234}) \,.
  \label{eq:Igamma-RGeq}
\end{align}

Next we derive the RG equation for the variation of the current rate $\delta\Gamma^\gamma(E)$.
We recall that the general RG equation for $\delta\Sigma_\gamma$ is obtained by taking the RG equation~\eqref{eq:dL-RGeq-general} for $\delta L(E)$ and replacing the leftmost coupling vertex of each diagram by the current vertex.
As in the RG equation for $\Gamma^\gamma$ only the $G^3$ part of the rightmost vertex can contribute since $\Sigma_\gamma\propto\tr$.
Using $\sum_{\sigma_1\sigma_2} \hat{L}^1_{\sigma_1\sigma_2}L^a\hat{L}^3_{\sigma_2\sigma_1} = \tr_\sigma \hat{L}^1 \hat{L}^3 = 6L^b$ we obtain
\begin{align}
    i\frac{\partial}{\partial E} \delta\Gamma^\gamma(E)
    &= -\frac32 \delta\hat{\bar\mu}_{12} I^\gamma_{12}(E) R(\hat{E}_{12}) G^3_{21}
    \notag \\ &\quad
    -i \frac34 I^\gamma_{12}(E) \Big[ R(\hat{E}_{12}) \delta\Gamma(\hat{E}_{12}) Z(\hat{E}_{12})
    \notag \\ &\quad\hphantom{+}
    + Z(\hat{E}_{12}) \delta\Gamma(\hat{E}_{12}) R(\hat{E}_{12}) \Big] G^3_{21}(\hat{E}_{12}) \,.
    \label{eq:dGammaL-RGeq}
  \end{align}
This is the only RG equation involving $\delta\Gamma(E)$, which appears in the form $\delta\Gamma\, O(J^3)$. It is therefore sufficient to know $\delta\Gamma(E)$ only in leading order $O(J)$ and we can neglect terms of the order $O(J^3/E)$ in the RG equation for $\partial_E\delta\Gamma(E)$.
The RG equation for $\partial_E \delta\Gamma(E)$ is
\begin{equation}
  -i\frac{\partial}{\partial E} \delta\Gamma(E)
  = \delta\hat{\bar\mu}_{12} G^2_{12}(E) R(\hat{E}_{12}) G^2_{21}(\hat{E}_{12}) \,.
\end{equation}

\subsection{Summary of the RG equations}
To bring the RG equations to the form of a closed set of first order ordinary differential equations we use $Z(E)^{-1}=1+i\partial_E \Gamma(E)$.
\begin{widetext}
  \begin{align}
    \frac{\partial}{\partial E} \Gamma(E)
    &= -i \left( \frac1{Z(E)} - 1 \right) \,, \label{eq:RGeq-gamma} \\
    \frac{\partial}{\partial E} Z(E)
    &= Z(E)\,G^2_{12}(E)\,\Ra(\hat{E}_{12})\,G^2_{21}(\hat{E}_{12})\,Z(E) \,, \\
    \frac{\partial}{\partial E} G^2_{12}(E)
    &=  \frac12 G^2_{13}(E)\,\Ra(\hat{E}_{13})\,G^2_{32}(\hat{E}_{13})
      + \frac12 G^2_{32}(E)\,\Ra(\hat{E}_{32})\,G^2_{13}(\hat{E}_{32}) \notag\\&\quad
      - \frac14 G^2_{34}(E) \left[ \Ra(\hat{E}_{34})\,G^2_{12}(\hat{E}_{34})\,Z(\hat{E}_{1234}) + Z(\hat{E}_{34})\,G^2_{12}(\hat{E}_{34})\,\Ra(\hat{E}_{1234}) \right] G^2_{43}(\hat{E}_{1234}) \,, \\
    \frac{\partial}{\partial E} G^3_{12}(E)
    &=  G^2_{13}(E)\,\Ra(\hat{E}_{13})\,G^3_{32}(\hat{E}_{13})
      + G^2_{32}(E)\,\Ra(\hat{E}_{32})\,G^3_{13}(\hat{E}_{32}) \notag\\&\quad
      + \frac12 G^2_{34}(E) \left[ \Ra(\hat{E}_{34})\,G^2_{12}(\hat{E}_{34})\,Z(\hat{E}_{1234}) + Z(\hat{E}_{34})\,G^2_{12}(\hat{E}_{34})\,\Ra(\hat{E}_{1234}) \right] G^3_{43}(\hat{E}_{1234}) \,, \\
    \frac{\partial}{\partial E} I^\gamma_{12}(E)
    &=  I^\gamma_{13}(E)\,\Ra(\hat{E}_{13})\,G^2_{32}(\hat{E}_{13})
      + I^\gamma_{32}(E)\,\Ra(\hat{E}_{32})\,G^2_{13}(\hat{E}_{32}) \notag\\&\quad
      + \frac12 I^\gamma_{34}(E) \left[ \Ra(\hat{E}_{34})\,G^2_{12}(\hat{E}_{34})\,Z(\hat{E}_{1234}) + Z(\hat{E}_{34})\,G^2_{12}(\hat{E}_{34})\,\Ra(\hat{E}_{1234}) \right] G^2_{43}(\hat{E}_{1234}) \,, \\
    \frac{\partial}{\partial E} \delta\Gamma(E)
    &= i \delta\hat{\bar\mu}_{12}\, G^2_{12}(E)\,\Ra(\hat{E}_{12})\,G^2_{21}(\hat{E}_{12}) \,, \\
    \frac{\partial}{\partial E} \delta\Gamma^\gamma(E)
    &= i\frac32 \delta\hat{\bar\mu}_{12}\, I_{12}^\gamma(E) \Ra(\hat{E}_{12})  G^3_{21}(\hat{E}_{12}) \notag\\&\quad
      -\frac34 I^\gamma_{12}(E) \left[ \Ra(\hat{E}_{12})\,\delta\Gamma(\hat{E}_{12})\,Z(\hat{E}_{12})
      + Z(\hat{E}_{12})\,\delta\Gamma(\hat{E}_{12})\,\Ra(\hat{E}_{12}) \right] G^3_{21}(\hat{E}_{12}) \,, \\
    \frac{\partial^2}{\partial E^2} \Gamma^\gamma(E)
    &= i \frac32 I^\gamma_{12}(E)\,\Ra(\hat{E}_{12})\,G^3_{21}(\hat{E}_{12})
\,. \label{eq:RGeq-last}
  \end{align}
\end{widetext}
These equations differ slightly from the zero temperature limit of the RG equations derived in Ref.~\cite{Reininghaus14} because here all quantities are Floquet matrices which do not commute and thereby prevent some simplifications.
Here the indices $1,\ldots$ only indicate reservoir labels, $1=\alpha=L,R$, and summation over equal indices is implicit.
All products are to be understood as Floquet matrix products and $\tfrac1Z$ denotes matrix inversion of the Floquet matrix $Z(E)_{nm}$.
When calculating the DC differential conductance directly from the RG equations we will fix $\gamma=L$ and $\delta\hat{\bar\mu}_{12}=(\delta_{1L}-\delta_{2L})\delta\vdc$ where $\delta\vdc$ is an infinitesimal parameter.

\subsection{Shifts in energy arguments}
\label{sec:energy-shifts-RGeq}
The RG equations at energy argument $E$ depend on $\Gamma$, $Z$, $G^2$ and $G^3$ at energy arguments $E+n\hat{\bar\mu}$ with $n=-2,-1,\ldots,2$. A numerical implementation needs to treat these quantities evaluated at different energy arguments separately. Hence we also need RG equations at energy argument $E+n\hat{\bar\mu}$ and the RG equations are not closed, but include an infinite number of coupled equations evaluated at different energy arguments.
However, the energy shift only enters as a next-to-leading order effect in the RG equations and can therefore be approximated.
The infinite set of coupled RG equations can be truncated by limiting the shift $n\hat{\bar\mu}$ to $|n|\leq n_\mathrm{max}$ and replacing, e.g., $\Gamma(E+[n_\mathrm{max}+1]\hat{\bar\mu})$ by $\Gamma(E+n_\mathrm{max}\hat{\bar\mu})$ in the RG equations. In the numerical evaluation $n_\mathrm{max}=3$ is sufficient to achieve convergence with only small deviations from larger $n_\mathrm{max}$.

\section{Initial conditions}
\label{sec:initial-conditions}
To find the initial conditions for the RG flow we use a two-step procedure.
First we use bare perturbation theory at $\im E = D \gg \tkv$ to compute bare and leading order effective quantities. Then we use these results to compute the RG flow in equilibrium and take the limit $D\to\infty$ analytically. This will allow us to construct initial conditions for the nonequilibrium RG flow at $E=i\Lambda_0$ for some $\Lambda_0\gg\tkv$.

The reason for this two-step procedure is that we cannot directly calculate initial conditions for $\Gamma(E)$ using bare perturbation theory.
However, we will see that a boundary condition for the RG flow of $\Gamma(E)$ can be found in equilibrium, enabling the construction of the full RG flow in equilibrium.
At $\im E\gg\tkv$ the equilibrium results with a perturbative treatment of the bias voltage provides a very good estimate for the quantities appearing in the RG equations in nonequilibrium. The high energy scale $E=i\Lambda_0$ and the high spin relaxation rate suppress all nonequilibrium effects.
Thus we can use the equilibrium RG flow to compute all quantities of interest at $E=i\Lambda_0$ and use these values as initial conditions for the nonequilibrium RG flow.

\subsection{Bare perturbation theory}

When we want to compute bare quantities at $\im E=D\gg\tkv$ it is sufficient to consider only leading order contributions in the bare coupling $J^{(0)}$, because $J^{(0)}$ vanishes in the limit $D\to\infty$ when keeping $\tkv$ fixed.
We have already seen in \Eq{eq:bare-vertex-leading-L2} that the bare vertex only contributes to the component $G^2$ of the effective vertex. To leading order in the bare coupling $G^2(E=iD)$ is thus given by the bare vertex,
\begin{equation}
  G^2_{11'}(E=iD) = -J_{\alpha\alpha'}^{(0)} = - 2\sqrt{x_\alpha x_{\alpha'}} J^{(0)} \,.
  \label{eq:G2-bare-initial-conditions}
\end{equation}
Similarly, we can write the bare current vertex [\Eq{eq:bare-current-vertex}] averaged over the Keldysh indices for $\eta=-\eta'=+$ in the form
\begin{equation}
  \Inoeta^{\gamma(0)}_{11'} = -\frac14 (\delta_{\alpha\gamma}-\delta_{\alpha'\gamma}) J_{\alpha\alpha'}^{(0)} \{\mathbf{S}\cdot\boldsymbol\sigma_{\sigma\sigma'}, \bullet\}
\end{equation}
to find the initial condition for its parametrization:
\begin{equation}
  I^\gamma_{11'}(E=iD) = (\delta_{\alpha\gamma}-\delta_{\alpha'\gamma}) 2\sqrt{x_\alpha x_{\alpha'}}J^{(0)} \,.
  \label{eq:Igamma-bare-initial-conditions}
\end{equation}
For $G^3$ there is no first order contribution and we need to compute the second order diagrams of the effective vertex:
\begin{align}
  &\Gfull^{p_1p_2}_{12}(E)-\Gfull^{(0)p_1p_2}_{12} = \diagram{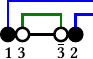} + \diagram{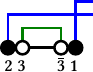} \\
  &= \int d\omega\, \gamma^{p_4}(\bar\omega_3) \Gfull^{(0)p_1p_3}_{13} \frac1{E+\bar\omega_3-\hat{\bar\mu}_{13}} \Gfull^{(0)p_4p_2}_{\bar32}
  \notag\\&\qquad
  - (1\leftrightarrow2) \\
  &= \Gfull^{(0)p_1p_3}_{13} \int_0^\infty d\omega\,D(\omega)\frac{p_4}{E-p_4\omega-\hat{\bar\mu}_{13}} \Gfull^{(0)p_4p_2}_{\bar32}
  \notag\\&\qquad
  - (1\leftrightarrow2) \\
  &= \Gfull^{(0)p_1p_3}_{13} \bigg\{ -i\frac{\pi}{2} p_4 + \log\Big(\frac{-iE+i\hat{\bar\mu}_{13}}{D}\Big)
  \notag\\&\qquad
  + O\Big[\frac{E}{D} \log\Big(\frac{-iE}{D}\Big)\Big] \bigg\} \Gfull^{(0)p_4p_2}_{\bar32} - (1\leftrightarrow2) \,.
  \label{eq:G3-bare-1}
\end{align}
Here $(1\leftrightarrow2)$ indicates the same expression with interchanged multi-indices $1,2$ and interchanged Keldysh indices $p_1,p_2$.
Since we are eventually only interested in the universal part of the diagram that remains relevant for $\im E \ll D$, but we evaluate it at $E=iD$, we can neglect the logarithmic term.
To calculate the initial condition for the parametrization of the coupling vertex it is sufficient to consider $\eta_1=-\eta_2=+$.
With this condition we can sum over the Keldysh indices in \Eq{eq:G3-bare-1} using the relation
\begin{align}
  \sum_{pp'} p \Gfull^{(0)pp'}_{13}
  &= \frac12 J_{\alpha_1\alpha_3}^{(0)} \{\mathbf{S}\cdot\boldsymbol\sigma_{\sigma_1\sigma_3}, \bullet\} \\
  &= \frac12 J_{\alpha_1\alpha_3}^{(0)} (\hat{L}^1 + \hat{L}^3)_{\sigma_1\sigma_3} \,.
  \label{eq:sump_pG}
\end{align}
By inserting this and \Eq{eq:bare-vertex-leading-L2} in \Eq{eq:G3-bare-1} and neglecting all logarithmic terms, we obtain the universal part of the coupling vertex to second order,
\begin{align}
  &\sum_{p_1,p_2} \big[ \Gfull^{p_1p_2}_{12}(E)-\Gfull^{(0)p_1p_2}_{12} \big] \notag\\\
  &= i\frac\pi4 J^{(0)}_{13} J^{(0)}_{32} \{ \hat{L}^2 (\hat{L}^1+\hat{L}^3) - [\hat{L}^{2\top} (\hat{L}^1+\hat{L}^3)^\top]^\top\} \\
  &= i2\pi \sqrt{x_1x_2} (J^{(0)})^2 \hat{L}^3 \,.
\end{align}
With this we can summarize the initial condition for $G^3$:
\begin{equation}
G^3_{12}(E=iD) = i\pi2\sqrt{x_1x_2}\,(J^{(0)})^2 \,.
  \label{eq:G3-bare-initial-conditions}
\end{equation}
Here we also note that there is no contribution to $G^a$ of order $O(J^2)$ in the initial conditions, even at finite voltage.

For the rate $\Gamma$, its derivative and variation, and for the current rate $\Gamma^\gamma$ one can calculate diagrams of the form $\diagram{L2}$ to obtain initial conditions. However, as we continue now we will see that we can obtain all required initial conditions without calculating this diagram.

\subsection{Equilibrium RG flow}
The results from bare perturbation theory will now allow us to calculate the equilibrium RG flow.
In equilibrium we do not need any Floquet matrices and can instead use scalar functions in Fourier space. Since the Floquet matrices are constructed as a generalization of Fourier space functions, we can use the RG equations that we derived for Floquet matrices also for scalar Fourier transforms.
After the derivation in Fourier space we will reformulate the results in Floquet space.

To simplify the equilibrium RG equations we use the ansatz
\begin{align}
  G^2_{12} &= -2\sqrt{x_1x_2} J \,, \label{eq:G2-parametrization-J} \\
  G^3_{12} &= 2\sqrt{x_1x_2} K \,, \label{eq:G3-parametrization-K} \\
  I^\gamma_{12} &= (\delta_{1\gamma} - \delta_{2\gamma}) M \label{eq:I-parametrization-M}
\end{align}
and note that $\delta\Gamma(E)=\Gamma^\gamma(E)=0$.
For $E=i\Lambda$ and a time-independent variation of the voltage $\delta\hat{\bar\mu}_{12} = (\delta_{1\gamma} - \delta_{2\gamma}) \delta\vdc$ the RG equations \eqref{eq:RGeq-gamma}--\eqref{eq:RGeq-last} simplify to~\cite{Reininghaus14}
\begin{align}
  \frac{d}{d\Lambda}\Gamma(\Lambda) &= \frac1{Z} - 1 \label{eq:RGeq-equilibrium-gamma-1} \,, \\
  \frac{d}{d\Lambda}Z(\Lambda) &= 4 \frac{Z^2 J^2}{\Lambda+\Gamma} \,, \\
  \frac{d}{d\Lambda}J(\Lambda) &= -2\frac{J^2(1+ZJ)}{\Lambda+\Gamma} \label{eq:RGeq-equilibrium-j-1} \,, \\
  \frac{d}{d\Lambda}K(\Lambda) &= -4\frac{JK(1-ZJ)}{\Lambda+\Gamma} \,, \\
  \frac{d}{d\Lambda}M(\Lambda) &= -2\frac{JM}{\Lambda+\Gamma} \,, \\
  \frac{d}{d\Lambda}\delta\Gamma^\gamma(\Lambda) &= i6\sqrt{x_Lx_R} \frac{MK}{\Lambda+\Gamma} \delta\vdc \,.
\end{align}
Like in Ref.~\cite{Reininghaus14} we define $\lambda=\Lambda+\Gamma$, $\tilde{J}=JZ$, $\tilde{M}=ZM$ and multiply by $\frac{d\Lambda}{d\lambda} = Z$ to obtain
\begin{align}
\frac{d}{d\lambda}Z &= 4 \frac{Z \tilde{J}^2}{\lambda} \,, \\
  \frac{d}{d\lambda}\tilde{J} &= -2\frac{\tilde{J}^2(1-\tilde{J})}{\lambda} \label{eq:RGeq-equilibrium-jtilde} \,, \\
  \frac{d}{d\lambda}K &= -4\frac{\tilde{J}K(1-\tilde{J})}{\lambda} \,, \\
  \frac{d}{d\lambda}\tilde{M} &= -2\frac{\tilde{J}\tilde{M}(1-2\tilde{J})}{\lambda} \,, \\
  \frac{d}{d\lambda}\delta\Gamma^\gamma &= i6\sqrt{x_Lx_R} \frac{\tilde{M}K}{\lambda} \delta\vdc \,.
\end{align}
Integrating \Eq{eq:RGeq-equilibrium-jtilde} yields the constant
\begin{equation}
  \tkrg \coloneqq \lambda \sqrt{\frac{\tilde{J}}{1-\tilde{J}}} e^{-\frac1{2\tilde{J}}} \label{eq:tkrg}
\end{equation}
which defines our reference energy scale as we will see below.
First we further simplify the equilibrium RG equations by multiplying with $\frac{d\lambda}{d\tilde{J}}$:
\begin{align}
  \frac{d}{d\tilde{J}} Z &= -2\frac{Z}{1-\tilde{J}} \,, \\
  \frac{d}{d\tilde{J}} K &= 2\frac{K}{\tilde{J}} \,, \\
  \frac{d}{d\tilde{J}} \tilde{M} &= \frac{\tilde{M}}{\tilde{J}} - \frac{\tilde{M}}{1-\tilde{J}} \,, \\
  \frac{d}{d\tilde{J}} \delta\Gamma^\gamma &= -i3\sqrt{x_Lx_R} \frac{\tilde{M}K}{\tilde{J}^2(1-\tilde{J})} \delta\vdc \,.
\end{align}
Integrating these equations yields
\begin{align}
  K &= K_0 \tilde{J}^2 \,, \label{eq:RGeq-equilibrium-result-K} \\
  Z &= Z_0 (1-\tilde{J})^2 \,, \\
  \tilde{M} &= M_0 \tilde{J}(1-\tilde{J}) \,, \\
  \delta\Gamma^\gamma &= -i\frac32\sqrt{x_Lx_R}M_0K_0 \tilde{J}^2 \delta\vdc \label{eq:RGeq-equilibrium-result-deltaGamma} \,.
\end{align}
In this analytic solution we can now insert the results obtained from bare perturbation theory for $\Lambda=D$. Taking the limit $D\to\infty$ we find that $K_0=i\pi$ [\Eq{eq:G3-bare-initial-conditions}] and $M_0=2\sqrt{x_Lx_R}$ [\Eq{eq:Igamma-bare-initial-conditions}].
To fix $Z_0$ we remember that when considering the current rate, terms scaling linearly with $E$ are nonuniversal and should not be included in the initial conditions.
Thus without nonuniversal terms we find $Z_0=1$.
At $\Lambda=0$ we know the equilibrium conductance $\delta\Gamma^\gamma/\delta\vdc=G_0=\frac{2e^2}{h} 4x_Lx_R$~\cite{Ng88} of the Kondo model and can thereby fix $\tilde{J}(0)=\frac{2}{\pi\sqrt3}$.
An equivalent condition can be obtained by using bare perturbation theory to calculate the current or differential conductance.
The only remaining initial condition for $\Gamma(0)$ has the unit of an energy and can therefore be taken as a reference energy scale.
In practice, when solving the RG equations we define the energy scale by the invariant $\tkrg$.

\subsection{Summary of the initial conditions for scalars}
\label{sec:initial-conditions-summary}
The procedure for obtaining initial conditions of the RG flow can be summarized as follows.
We start at $E=i\Lambda=0$ with the initial conditions
\begin{align}
  \tilde{J}(0) &= \frac{2}{\pi \sqrt3} \label{eq:init-conditions-E0-J} \\
  \Gamma(0) &= \tkrg \sqrt{\tfrac{1-\tilde{J}(0)}{\tilde{J}(0)}} e^{\frac1{2\tilde{J}(0)}} \\
  Z(0) &= (1-\tilde{J})^2 \label{eq:init-conditions-E0-Z}
\end{align}
where $\tkrg$ defines the energy scale and can be set to $1$.
Then we numerically integrate the RG equations \eqref{eq:RGeq-equilibrium-gamma-1}--\eqref{eq:RGeq-equilibrium-j-1} from $\Lambda=0$ to $\Lambda_0$ where $\Lambda_0\gg\tkrg$ is a parameter of the numerical solution. For all data presented in this paper we use $\Lambda_0=10^9 \tkrg$, which is sufficient to achieve numerical convergence.
The resulting values for $\tilde{J}(\Lambda_0)=J(\Lambda_0)Z(\Lambda_0)$ and $Z(\Lambda_0)=[1-\tilde{J}(\Lambda_0)]^2$ allow us to construct the remaining initial conditions by following Eqs.~\eqref{eq:RGeq-equilibrium-result-K}--\eqref{eq:RGeq-equilibrium-result-deltaGamma}:
\begin{align}
  G^2_{12}(i\Lambda_0) &= -2\sqrt{x_1x_2} J(\Lambda_0) \label{eq:init-conditions-final-g2} \\
  G^3_{12}(i\Lambda_0) &= i\pi 2\sqrt{x_1x_2}\tilde{J}(\Lambda_0)^2 \\
  I^\gamma_{12}(i\Lambda_0) &= (\delta_{1\gamma}-\delta_{2\gamma}) 2\sqrt{x_Lx_R} J(\Lambda_0)[1-\tilde{J}(\Lambda_0)] \\
  \delta\Gamma^\gamma(i\Lambda_0) &= 3\pi x_L x_R \tilde{J}(\Lambda_0)^2 \delta\vdc \,.
\end{align}
From the equilibrium RG equations we furthermore find that $\delta\Gamma(i\Lambda)=0$ and $\partial_\Lambda \delta\Gamma^\gamma(i\Lambda)/\delta\vdc$ is of the order of $J^3/\Lambda \ll \tkrg^{-1}$.
This yields the initial conditions
\begin{align}
  \delta\Gamma(i\Lambda_0) &= 0 \,, \\
\frac{\partial\Gamma^\gamma}{\partial E}(i\Lambda_0) &= 0 \,. \label{eq:init-conditions-final-y}
\end{align}
The initial condition for $\Gamma^\gamma(i\Lambda_0)$ will be discussed in the next section.

These initial conditions are constructed such that the direct calculation of the differential conductance $G$ via $\delta\Gamma^\gamma$ will yield the universal limit $G=2e^2/h$ exactly for $V(t)\equiv0$, independent of the cutoff $\Lambda_0$. The same result can be obtained when using the more complicated bare perturbation result for the current to derive initial conditions for $\Gamma^\gamma$ and $\delta\Gamma^\gamma$.
Taking the whole procedure of the FRTRG with $\Lambda_0$ as a parameter we find that this construction leads to a good convergence of $\delta\Gamma^\gamma(E=0)$ for large $\Lambda_0$ also in nonequilibrium and a slightly slower convergence of the current rate $\Gamma^\gamma(E=0)$.
The comparison of the differential conductance computed from $\delta\Gamma^\gamma$ and from the derivative of the current constructed via $\Gamma^\gamma$ (provided in App.~\ref{app:g-self-consistency}) shows uncertainties of less than $4\%$ for our choice of parameters of the numerical evaluation.

\subsection{Energy shifts and Floquet matrices}
\label{sec:energy-shifts-initial-conditions}
Now that we have initial conditions in equilibrium we still need to transform them to Floquet space and include the energy shifts explained in \autoref{sec:energy-shifts-RGeq}.
Using \Eq{eq:fourier-floquet} we immediately find that without any energy shift the Floquet matrix $\hat{\Gamma}(E)$ obtained from the scalar function $\Gamma(E)$ in equilibrium is given by $\hat{\Gamma}(E)=\Gamma(E+\Omega\hat{N})$.
Since the argument of the function $\Gamma$ is a diagonal matrix, we can simply act element-wise with the function $\Gamma$ on the diagonal entries of this matrix.
Thus we need to compute $\Gamma(i\Lambda_0+n\Omega)$ for integer values of $n$. This is possible because the equilibrium RG flow allows us to reach arbitrary scalar energy arguments $E$ of $\Gamma(E)$ by numerically solving ordinary differential equations.

Moreover, from \autoref{sec:energy-shifts-RGeq} we remember that the RG equations include the Floquet matrices $\Gamma$, $Z$, $G^2$, and $G^3$ not only at a scalar energy $E$, but also at $E+n\hat{\bar\mu}$ with $n$ denoting an integer and $\hat{\bar\mu}$ a non-diagonal Floquet matrix.
For the initial conditions this implies that we need to evaluate $\Gamma(E)$ also with a non-diagonal Floquet matrix as energy argument to obtain the initial condition
$\hat\Gamma(E+n\hat{\bar\mu})=\Gamma(E+\Omega\hat{N}+n\hat{\bar\mu})$
at $E=i\Lambda_0$. To do so, we diagonalize the argument of the function $\Gamma$ and evaluate $\Gamma$ at the eigenvalues using again the equilibrium RG flow.

The same procedure as explained above for $\Gamma$ also applies to the expressions which we obtained for the initial conditions for $Z$, $G^2$ and $G^3$. For the other quantities appearing in the RG equations we do not need to include any shifts by $\hat{\bar\mu}$, such that for these other quantities the initial conditions are given by diagonal Floquet matrices.

A special case and exception of the previous statement is the initial condition for the current rate $\Gamma^\gamma$.
Since we have initial conditions for the differential conductance $\delta\Gamma^\gamma$ and can treat the voltage perturbatively at $E=i\Lambda_0$, the initial condition for the current is given by
\begin{equation}
  \hat{\Gamma}^L(i\Lambda_0) = \hat{\bar\mu}_{LR} \frac{\delta\Gamma^L(i\Lambda_0+\Omega\hat{N})}{\delta\vdc} \,.
  \label{eq:gammaL-initial-condition-final}
\end{equation}
In this equation it is important to note that $\delta\Gamma^L(i\Lambda_0+\Omega\hat{N})$ is a diagonal Floquet matrix and the product of the two Floquet matrices on the right hand side makes sure that column $n$ of Floquet matrix $\Gamma^\gamma$ obtains the frequency shift $n\Omega$ in agreement with the definition of the Floquet matrices in \Eq{eq:floquet-matrix}.

\section{Summary of the method}
\label{sec:method-summary}
The procedure to numerically calculate physical observables can be summarized as follows.
Starting from the initial conditions Eqs.~\eqref{eq:init-conditions-E0-J}--\eqref{eq:init-conditions-E0-Z} at $E=0$ we numerically integrate the equilibrium RG flow equations \eqref{eq:RGeq-equilibrium-gamma-1}--\eqref{eq:RGeq-equilibrium-j-1} until we reach $E=i\Lambda_0$ where $\Lambda_0\sim 10^8\tkv$ is some large number.
The results of the equilibrium RG flow at $E=i\Lambda_0$ are converted to the notation used in the nonequilibrium RG flow using Eqs.~\eqref{eq:init-conditions-final-g2}--\eqref{eq:gammaL-initial-condition-final}.
To obtain the initial conditions for the nonequilibrium RG flow at $E=i\Lambda_0$ we furthermore include Floquet-matrix-valued shifts of the energy argument as explained in \autoref{sec:energy-shifts-initial-conditions}.
The nonequilibrium RG equations \eqref{eq:RGeq-gamma}--\eqref{eq:RGeq-last} can then be integrated numerically from $E=i\Lambda_0$ to $0$.
Eventually the results of the nonequilibrium RG flow at $E=0$ are used to compute observables such as the current.

\subsection{Calculating observables}
By solving the RG equations we obtain the Floquet matrix representation of the current rate $\Gamma^\gamma$ and its variation $\delta\Gamma^\gamma$ at frequency $E=0$.
These Floquet matrices contain the Fourier series of the current and the differential conductance, as we will see next.
When calculating observables we are only interested in the case that the evolution started a long time ago, $t_0\to-\infty$.
The definition of the Floquet matrix in Eqs.~\eqref{eq:floquet_1} and \eqref{eq:floquet-matrix} implies that for $t_0=-\infty$ the central column of the Floquet matrix of the current rate $\Gamma^\gamma(E)$ is the Fourier series of
\begin{equation}
  \sum_n e^{-in\Omega t} \Gamma^\gamma(E)_{n0}
  = \int_{-\infty}^t ds\, e^{iE(t-s)} \Gamma^\gamma(t, s) .
  \label{eq:current-fourier-rep-1}
\end{equation}
At $E=0$ we can identify the right hand side of \Eq{eq:current-fourier-rep-1} as the current $I(t)$ through the system, because
\begin{align}
  I(t)
  &= -i\int_{-\infty}^t \!ds\, \tr \Sigma_\gamma(t,s) \rho(s) \\
  &= \int_{-\infty}^t \!ds\, \Gamma^\gamma(t,s) \tr\rho(s) \,.
\end{align}
Here we dropped the index $\gamma$ for the current to avoid confusion with the current vertex.
Thus we can identify the Fourier modes of the current $I(t)$ with matrix elements of the Floquet matrix $\Gamma^\gamma(E=0)$,
\begin{equation}
  I_n = \Gamma^\gamma(0)_{n0} \,.
\end{equation}
In analogy to the current we can calculate the differential conductance $G(t) = dI(t)/d\vdc = \sum_n e^{-in\Omega t} \delta\Gamma^\gamma(0)_{n0}/\delta\vdc$.
When reaching $E=0$ in the RG flow we thus know the full time dependence of the current and the differential conductance, which are computed independent of each other.

 \section{Results}
\label{sec:results}
The FRTRG allows us to compute the current as a function of time for an arbitrary periodic bias voltage.
We will first focus on the case of harmonic driving and discuss the conductance of the system, which can be compared to other predictions and experimental data.
Afterwards we will analyze the time dependence of the response current for different driving profiles.
In the following we consider only the case of symmetric coupling to the reservoirs, $x_L=x_R=1/2$, except if $x_L$ is explicitly given.

\subsection{Differential conductance}
\label{sec:macromotion}

\begin{figure}[t]
  \centering
   \includegraphics{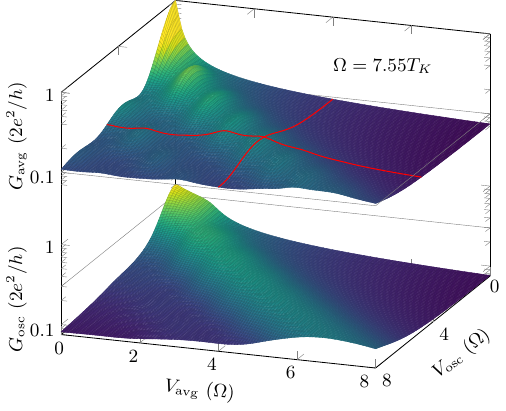}
   \caption{\label{fig:G-vdc-vac}Differential conductance $\gdc=d\idc/d\vdc$ and $\gac=d\iac/d\vac$ for $V(t)=\vdc+\vac\cos(\Omega t)$ as function of $\vdc$ and $\vac$ at $\Omega=7.55\tkv$.
    Red lines indicate the parameters which are compared to approximations in \autoref{fig:G-comparison}(a)--(b).
    The time-averaged differential conductance (upper panel) shows peaks at $\vdc=n\Omega$ weighted by $J_n(\vac/\Omega)^2$. The width of these peaks in $\gdc(\vdc)$ at constant $\vac$ is of the order of $\tkv$ while all other structures are smooth on the energy scale of $\Omega$.
    $\gac$ at constant $\vdc$ or $\vac$ shows a broad maximum at $\vdc\approx\vac$.
    This result can be understood from the low-frequency limit, where $\vdc=\vac$ implies that $V(t)\approx0$ for a relatively long time in each period, which leads to a high differential conductance due to the Kondo resonance.
  }
\end{figure}

We first consider the harmonic bias voltage
\begin{equation}
  V(t) = \vdc + \vac \cos(\Omega t)
  \label{eq:v-harmonic}
\end{equation}
and focus only on the zeroth and first Fourier mode of the current, postponing the discussion of higher harmonics to \autoref{sec:micromotion}.
In analogy to \Eq{eq:v-harmonic} we use the notation $I_\mathrm{avg}=I_0$ and $I_\mathrm{osc}=2|I_1|$ for these Fourier modes.
To characterize the electronic properties of the quantum dot we focus on the differential conductance for the average and oscillating voltage, $\gdc=d\idc/d\vdc$ and $\gac=d\iac/d\vac$.
While $\gac$ can only be computed by discrete differentiation of the current, the data presented here for $\gdc$ are calculated directly from the variation $\delta\vdc$ of $\vdc$ in the RG equations.
We use the convention that the Kondo temperature is defined by the voltage at which $\gdc$ drops to half of its universal value, $\gdc(\vdc=\tkv)=\gdc(\vdc=0)/2=e^2/h$ at $\vac=0$. This Kondo temperature is related to the energy scale used in the initial conditions by $\tkv=3.31\tkrg$.

An overview of $\gdc$ and $\gac$ for fixed frequency is given in \autoref{fig:G-vdc-vac}.
It is of no surprise that any finite bias voltage will perturb or destroy the equilibrium state and thereby reduce the Kondo resonance in the conductance. This can be observed in both $\gdc$ and $\gac$.
Besides this expected suppression, the conductance for the oscillating part of the current, $\gac$, is largest when $\vac\approx\vdc$, as one would expect in the adiabatic limit: For $\vdc=\vac$ and very slow driving the Kondo peak in the conductance at $V(t)=0$ enhances both $\gdc$ and $\gac$. The FRTRG results show this effect in the form of a broadened maximum of $\gac$ at $\vdc\approx\vac$ also at relatively fast driving.

The differential conductance for the average current, $\gdc$, shows a richer structure than $\gac$ and has been discussed before in the literature \cite{Hettler95,Ng96,Goldin98,Lopez98,Kaminski99,Kaminski00,Kogan04,Bruhat18}.
The suppression of the Kondo resonance in $\gdc(\vdc,\vac)$ at $\vdc=0$ as seen in \autoref{fig:G-vdc-vac} is particularly interesting because it originates from two different effects.
The driving leads to decoherence which suppresses the Kondo effect~\cite{Kaminski00}, but it also induces photon sidebands in the density of states and redistributes the weight of the central Kondo resonance to these sidebands~\cite{Hettler95}. Instead of just one Kondo resonance at the Fermi energy, the coherently driven system can have multiple copies of the Kondo resonance shifted by multiples of the driving frequency $\Omega$. These sidebands can be seen as satellite peaks in $\gdc$ at $\vdc=n\Omega$, $n\in\mathbb{Z}$.
Our results show that for $\Omega\gtrsim\tkv$ the satellite peaks will appear before decoherence becomes strong enough to completely blur out this structure.
This motivates an analysis comparing our full RG results with those neglecting the additional decoherence induced by the oscillating part of the voltage.

\begin{figure}[t]
  \centering
   \includegraphics{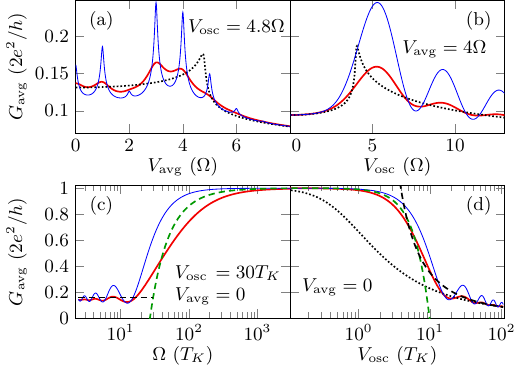}
   \caption{\label{fig:G-comparison}Differential conductance $\gdc(\vdc,\vac,\Omega)$ from FRTRG (red line)
    compared to the adiabatic approximation (black dotted line), the phenomenological \Eq{eq:G_approx} (thin blue line), and the analytic approximation for limiting cases from Ref.~\cite{Kaminski00} (dashed lines in (c)--(d)).
    We use $\Omega=7.55\tkv$ as in \autoref{fig:G-vdc-vac} except in panel (c).
Many features of these curves are due to the replicas of the Kondo resonance and are overestimated by the phenomenological approximation.
    Oscillations of $\gdc$ as function of $\Omega$ and $\vac$ at intermediate frequencies and high $\vac$ are not captured at all by the analytic approximations of Ref.~\cite{Kaminski00}.
    As expected, the adiabatic approximation (defined by setting $\Omega\to0$) is only applicable if $\Omega\ll\max\{\vac,\vdc,\tkv\}$.
  }
\end{figure}

We first estimate $\gdc$ by including the time-dependent part of the bias voltage only through photon-assisted tunneling. To this end we only consider the effect of $\vac$ on the density of states, while ignoring decoherence effects due to the oscillating voltage.
For simplicity let us assume that the voltage is applied to the chemical potential $\mu_L(t)$ of the left reservoir while $\mu_R$ is constant.
The voltage modulation effectively splits the energy levels in the left reservoir, leading to the effective density of states $\varrho'(E) = \sum_n J_n^2(\vac/\Omega) \varrho(E+n\Omega)$~\cite{Tien63},
which induces satellite peaks of the Kondo resonance in the equilibrium density of states $\varrho(E)$ weighted by Bessel functions.
The differential conductance for this effective density of states,
\begin{equation}
  \gdc(\vdc,\vac,\Omega)\approx\sum_{n=-\infty}^\infty J_n^2\left(\tfrac{\vac}{\Omega}\right) G_\mathrm{static}(\vdc+n\Omega)
  \,,
  \label{eq:G_approx}
\end{equation}
can be calculated by using the differential conductance $G_\mathrm{static}(V)=\gdc(\vdc=V,\vac=0)$ for the same system with a constant bias voltage~\cite{Goldin98}, which we take from an RTRG calculation.
In \autoref{fig:G-comparison} this approximation is shown as blue lines and can be compared to the full FRTRG results (red lines).
The comparison shows that this picture of photon assisted tunneling explains not only the satellite peak positions at $\vdc=n\Omega$, but it also predicts the relative peak weight $J_n^2(\vac/\Omega)$ in good agreement with the FRTRG results.
In \autoref{fig:G-comparison}(a) the satellite peaks and their weights are visible, and in \autoref{fig:G-comparison}(b) the broad maxima and minima are due to $J_n^2(\vac/\Omega)$ with $n=-\vdc/\Omega$ as the dominant term in \Eq{eq:G_approx}.

In contrast to the previously explained approximation, the FRTRG self-consistently takes into account processes causing decoherence of the many-body state.
Thus the Kondo resonance and its replicas described by \Eq{eq:G_approx} are partially suppressed and broadened.
However, as illustrated in \autoref{fig:G-vdc-vac} and \autoref{fig:G-comparison}(a), the decoherence is not strong enough to completely suppress the satellite peaks in the case of strong and sufficiently fast driving ($\vac>\Omega>\tkv$).
This key result is confirmed by the experiments of Refs.~\cite{Kogan04,Bruhat18} (compare also Fig.~\ref{fig:experiment}).
The decoherence due to the oscillating voltage reduces the height of the peaks in $\gdc(\vdc)$ while increasing the peak widths, as one can see in \autoref{fig:G-comparison}(a).
The total peak weight is unaffected by the decoherence because the average current $I(\vdc')=\int_0^{\vdc'} \gdc d\vdc$ at large, constant bias voltage $\vdc'\gg\vac$ must be independent of $\vac$.
These thoughts let us conclude that all effects seen in \autoref{fig:G-comparison}(a) can be qualitatively understood, while a quantitative description is achieved only by the here introduced FRTRG approach.

Generalizing the previous discussion to the full parameter dependence of $\gdc(\vdc,\vac,\Omega)$, one finds that many features can be qualitatively understood using the phenomenological approximation~\eq{eq:G_approx}. This approximation qualitatively describes the differential conductance in the whole parameter space, but generally overestimates
all effects related to the satellite resonances in the effective density of states
as illustrated in \autoref{fig:G-comparison}.
The FRTRG results for the differential conductance are consistent with known limiting cases like the adiabatic limit and the high frequency limit.
For $\vdc=0$ our results can be compared to the analytic predictions for limiting cases of Ref.~\cite{Kaminski00}, represented by dashed lines in \autoref{fig:G-comparison}(c)--(d)). Here the FRTRG describes the full crossover from slow to fast and weak to strong driving.
Besides an overall good agreement this comparison shows that beyond the prediction of Ref.~\cite{Kaminski00} the FRTRG data include partially suppressed resonance effects of the phenomenological approximation \Eq{eq:G_approx}.
The FRTRG achieves this by self-consistently including self-energy insertions using the RG flow.
This analysis highlights that to quantitatively describe the interplay of coherent driving and decoherence in the full crossover regime from weak to strong driving a theoretical approach like the one presented here is vital.

\subsection{Comparison to an experiment}
\label{sec:experiment}

\begin{figure}[t]
   \includegraphics{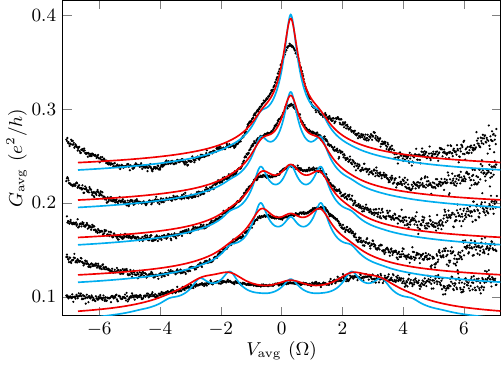}
   \caption{\label{fig:experiment}Comparison of FRTRG data (red and cyan lines) to measured differential conductance $\gdc=d\idc/d\vdc$ through a single electron transistor in the Kondo regime from Ref.~\cite{Kogan04} (black dots) at $\Omega/2\pi=13.47\,\mathrm{GHz}$ and $\tkv\approx0.5\Omega$.
    The red curves are calculated for symmetric coupling to the reservoirs ($x_L=x_R$) and rescaled by a factor $0.114$ to include finite temperature and asymmetry effects, whereas the cyan lines are obtained for an asymmetry of $x_L=0.05$ in the coupling and are only rescaled by a factor $0.62$ to account for finite temperature.
    The average voltage $\vdc$ in the theory curves is shifted to match the position of the central Kondo peak in the experimental data.
    The FRTRG data for asymmetric coupling show more pronounced peaks which are, however, expected to be smeared out by the finite temperature.
The curves show $\Omega=2.11\tkv$ and (starting from the top) $\vac=0.73\Omega, 1.13\Omega, 1.51\Omega, 1.68\Omega$, and $3.62\Omega$.
We adapted the calibration of $\vac$ because the procedure of calibrating $\vac$ in Ref.~\cite{Kogan04} is expected to be biased (cf.~App.~\ref{app:experiment}).
    The scale of the $y$ axis corresponds to the curve at the bottom and subsequent curves are separated by $0.04\,e^2/h$ for clarity.
}
\end{figure}

The satellite peaks in $\gdc(\vdc, \vac=constant)$ were measured by Kogan {\it et al.} \cite{Kogan04} and we compare our results to their experiment in \autoref{fig:experiment}.
The measured differential conductance through a single-electron transistor (black dots) and the FRTRG data (red curves) show good quantitative agreement in the regime of small bias voltage which is dominated by the Kondo physics,
presuming that the following considerations are taken into account.

Since our method only describes the universal limit of the Kondo model, a constant offset in the conductance is expected which we manually adapt when comparing theory and experiment.
The small but finite temperature of approximately $\tfrac13 \tkv$ in the experiment reduces the height of the conductance peaks and introduces another source of decoherence.
Also an asymmetric coupling to the two reservoirs may further reduce the overall conductance.
In \autoref{fig:experiment} the red lines show $0.055 e^2/h + 0.114 \gdc^\mathrm{theory}$, accounting for shift and rescaling due to the previously mentioned effects.
Besides this manual adjustment of the scale of $\gdc$, it is also possible to directly include an asymmetric coupling when solving the RG equations to explain the reduced overall conductance. Assuming that the prefactor of $\gdc^\mathrm{theory}$ can be explained entirely by finite temperature and asymmetric coupling, we find the cyan lines in \autoref{fig:experiment}. For this we estimate based on Ref.~\cite{Pletyukhov12a} that the finite temperature of $\tfrac13\tkv$ reduces the conductance by a factor $0.62$ and we use the asymmetry factor $x_L=0.05$ to further reduce $\gdc$.
Notably, the asymmetry leads to more pronounced peaks in the FRTRG prediction for $\gdc(\vdc,\vac=constant)$ at zero temperature, but such sharp peaks are not visible in the experiment. This apparent deviation might stem from our simplified treatment of the finite temperature.

Furthermore, the curves of theory and experiment only agree if the calibration of the oscillating part of the voltage is modified. In the experiment $\vac$ is gauged by comparison of the measured conductance $\gdc(\vdc,\vac)$ to the adiabatic limit at high $\tkv$.
However, a simulation of this procedure using FRTRG (provided in App.~\ref{app:experiment}) suggests that this procedure will underestimate $\vac$ because of deviations from the adiabatic limit.
We therefore multiply the values for $\vac$ from the experiment by a correction factor $1.4$ before comparing with our calculations.
Notably, this factor must be chosen larger than the rough estimate based on the simulated calibration procedure in App.~\ref{app:experiment}.
Bearing in mind that the necessity of shifting and rescaling $\gdc$ is not unexpected and that the modified calibration of $\vac$ can be explained by modeling the calibration procedure using FRTRG, we conclude that the FRTRG predictions show very good agreement with the experiment.

\subsection{Micromotion}
\label{sec:micromotion}

We now focus on the time dependence of the current within one driving period---the micromotion---which offers direct insights into the memory of the system.
We start by considering the case of harmonic driving with $\vdc=0$, before we analyze specific memory effects in detail by using a pulsed bias voltage.

\begin{figure}[t]
  \centering
   \includegraphics{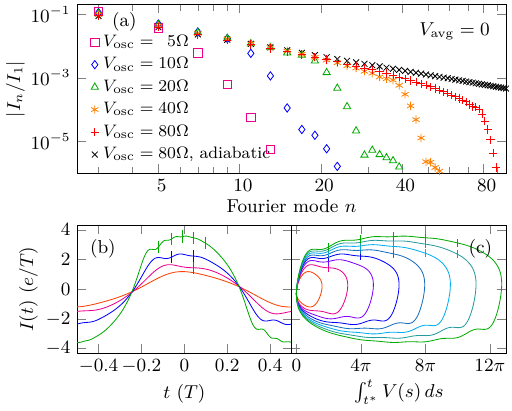}
   \caption{\label{fig:higher-harmonics}Current in time and Fourier domain for $V(t)=\vac\cos(\Omega t)$, $\Omega=7.55\tkv$.
    (a)~Higher harmonics in the response current relative to the first harmonic.
    Even modes vanish at $\vdc=0$ due to symmetry.
    The decay is approximately algebraic for $n<\vac/\Omega$ in agreement with the adiabatic limit $\Omega\to0$ (black), except for very high harmonics ($n\gtrsim 40$ for $\vac=80\Omega$). For $n>\vac/\Omega$ the modes decay exponentially. (b)~Time-resolved current for $\vac=2.5\Omega,5\Omega,10\Omega$, and $20\Omega$ (from red to green).
    Markers indicate times $t$ at which $\int_{t^*}^t V(s)\,ds$
is an integer multiple of $2\pi$.
    The reference time $t^*=-T/4$ is chosen such that $V(t^*)=0$.
    (c)~$I(t)$ versus $\int_{t^*}^t V(s)\,ds$ parametrized by $t$ for $\vac=2.5\Omega\ldots20\Omega$ in steps of $2.5\Omega$.
    The coherent oscillations in this representation reveal the origin of the oscillations in $I(t)$ as explained in the text.
}
\end{figure}

In \autoref{fig:higher-harmonics}(a)--(b) the current and its Fourier modes are plotted for strong and fast harmonic driving.
The Fourier modes may serve as an experimentally accessible observable. Though as expected we find a good agreement between FRTRG and the adiabatic limit on long time scales (small mode index~$n$), for $n\Omega\gtrsim\vac$ the Fourier modes drop exponentially to negligible values, which is not captured by the adiabatic limit.
Compared to the adiabatic limit this indicates a smoothening of the current on the time scale $1/\vac$ that corresponds to the largest energy scale in the system.
In time domain one can see further deviations from the adiabatic limit. Small oscillations in the current in \autoref{fig:higher-harmonics}(b) resemble the \enquote{ringing} structure that was predicted in Ref.~\cite{Jauho94} for non-interacting systems.
To understand this structure of the micromotion in time domain we will first discuss a different driving profile, which allows us to understand the underlying physical effects in a more refined setup. This will eventually provide an explanation for the structures in \autoref{fig:higher-harmonics}(b)--(c).

\begin{figure}[t]
  \centering
   \includegraphics{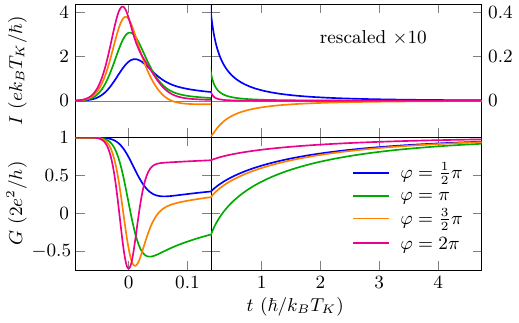}
   \caption{\label{fig:time-scales}Current and differential conductance as function of time for a Gaussian voltage pulse centered at $t=0$ of duration (full width at half maximum) $t_\mathrm{pulse}=0.058/\tkv$.
    For $\varphi=2\pi$ or $\pi$ the current quickly decays to zero after the pulse, while for $\varphi=\pi/2$ or $3\pi/2$ a slower decaying current persists after the pulse. For $\varphi\approx3\pi/2$ this current after the pulse is negative.
    The axes are chosen such that the areas under the curves for $I(t)$ can be directly compared between the left and the right panel.
The differential conductance $G(t)=dI(t)/d\vdc$ indicates the recovery of the Kondo resonance, which happens on a time scale of the order of $1/\tkv$, slightly slower than the decay of the current.
  }
\end{figure}

To see how the voltage affects the current in the Kondo model on short time scales, we now consider a single short voltage pulse applied to a system which is otherwise in equilibrium.
In order to solve this problem with FRTRG we need to repeat the pulse with a low frequency $\Omega<\tkv$ such that the voltage is time periodic but the repeated pulses do not affect each other and we can safely treat them as independent pulses.
Again we assume for simplicity that the voltage $V(t)$ is only applied to the chemical potential $\mu_L(t)$ of the left reservoir while the right reservoir is kept at constant chemical potential.
The voltage pulse causes a phase shift of the quantum state by $e^{i\varphi}$, $\varphi=\int V(t) dt$, for each electron that is in the left reservoir during the pulse.
To understand the relevance of this phase shift we consider an electron that can hop between the reservoirs before or after the voltage pulse. Depending on the reservoir in which the electron is during the pulse, it will either obtain the phase $e^{i\varphi}$ or it will obtain no extra phase. The interference of these two options---the electron hopping before or after the pulse---will cause oscillations of the current as a function of $\varphi$.
As we will see below, this type of interference effects is essential to understand the current on short time scales for strong and fast driving.

The current and differential conductance for a short Gaussian voltage pulse calculated using FRTRG are shown in \autoref{fig:time-scales} and both quantities confirm the relevance of $\varphi$ for the response after the voltage pulse.
The differential conductance $G(t)=dI(t)/d\vdc$ serves as an indicator for the destruction and recovery of the Kondo resonance.
For $\varphi=2\pi$ the differential conductance reaches a relatively high value immediately after the pulse, indicating that the Kondo resonance is mostly restored after a very short time.
This matches our expectation since for $\varphi=2\pi$ the interference of states with or without the phase shift $e^{i\varphi}$ will show no deviation from the equilibrium case $\varphi=0$.
In contrast, for $\varphi=\pi$ the differential conductance only slowly approaches its equilibrium value after being pushed to $G(t)<0$ by the short pulse.
In this case the maximal deviation of $e^{i\varphi}$ from its equilibrium value comes with a strong suppression of the Kondo resonance.
For $\varphi=\pi/2$ and $\varphi=3\pi/2$ \autoref{fig:time-scales} shows two very similar curves of $G(t)$ after the pulse. Since $G(t)$ does not depend on the sign of the bias voltage, we can interpret this as the equivalence of $\varphi=3\pi/2$ and $\varphi=-\pi/2$, which confirms that the effect of the voltage pulse on the Kondo resonance can be characterized by $e^{i\varphi}$.

A simple interpretation of these effects focuses on the limited memory of the system. The Kondo system contains memory only in the system-reservoir correlations. As we have seen before, electrons that can hop before or after the pulse cause interference depending on $e^{i\varphi}$ and must therefore include information about $e^{i\varphi}$ in the system memory. Electrons hopping during the voltage pulse can add other information to the memory, but their contribution to the memory is small for very short pulses. For very short pulses of duration $t_\mathrm{pulse}\ll 1/\tkv$ one can therefore approximate that after the pulse only information about $e^{i\varphi}$ remains in the system memory, such that only this phase can influence the current and the differential conductance after the pulse.

This interpretation is also supported by the current.
\autoref{fig:time-scales} indicates that the charge transported in response to a short voltage pulse consists of two contributions: electrons tunneling during the voltage pulse that are immediately affected by the voltage, and a retarded current of electrons tunneling after the pulse.
The latter contribution stems from the previously discussed electron hopping processes and relies on the memory of the system.
Since for short pulses this memory mainly contains information about $e^{i\varphi}$, we expect also here that $\varphi=3\pi/2$ and $\varphi=-\pi/2$ lead to approximately the same retarded current.
This implies that for $\varphi=3\pi/2$ the current after the pulse will flow in reverse direction, because it will have the same sign as the current caused by a negative voltage pulse with $\varphi=-\pi/2$.
The pulse with $\varphi=3\pi/2$ therefore causes a \emph{counter-propagating} current after the pulse as we see in \autoref{fig:time-scales}.
A counter-propagating current after the pulse is expected more generally for short pulses with $\pi<\varphi<2\pi$ and, albeit suppressed, for $\varphi\approx 3\pi/2 + 2n\pi$, $n=1,2,\ldots$.
\autoref{fig:time-scales} also shows a quickly decaying current for $\varphi=\pi$ and $2\pi$.
In these cases the memory---in the approximation for very short pulses---does not contain information about the sign of the voltage pulse such that the current must decay very quickly.

\begin{figure}[t]
  \centering
   \includegraphics{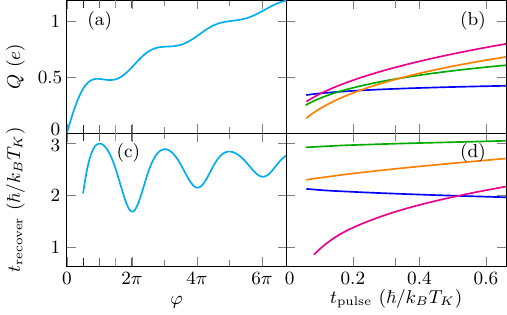}
   \caption{\label{fig:pulse-interference}Relevance of the phase difference $\varphi=\int dt\, V(t)$.
    (a)~Average charge $Q$ transported by a single Gaussian voltage pulse of constant duration $t_\mathrm{pulse}=0.33/\tkv$ and variable height.
    (b)~Charge $Q$ for short pulses of constant phase $\varphi=\pi/2,\pi,3\pi/2$ and $2\pi$ (blue, green, orange and magenta).
    $t_\mathrm{pulse}$ is the full width at half maximum of a Gaussian pulse.
    (c)--(d)~Time $t_\mathrm{recover}$ needed to recover the Kondo resonance after a Gaussian voltage pulse for the same $V(t)$ as in (a)--(b).
    The effect of $\varphi$ on the recovery time clearly increases for shorter pulses.
  }
\end{figure}

The strong dependence of the retarded current on $e^{i\varphi}$ can also be seen in the total transported charge $Q$, which oscillates as function of $\varphi$ as shown in \autoref{fig:pulse-interference}(a)--(b).
In \autoref{fig:pulse-interference}(b) we see that for shorter pulses the contribution of the retarded current becomes more and more relevant, such that a relatively weak pulse with $\varphi=\pi/2$ can transport more charges than a strong pulse with $\varphi=3\pi/2$.
Also the destruction and recovery of the Kondo resonance depends on both the phase and the pulse duration.
To quantify the destruction of the Kondo resonance we define $t_\mathrm{recover}$ as the time needed after the pulse until $G(t)$ reaches 80\% of the equilibrium value, $G(t_\mathrm{recover})=1.6e^2/h$.
In \autoref{fig:pulse-interference}(c) one can see oscillations of $t_\mathrm{recover}$ as function of $\varphi$ for short Gaussian voltage pulses of different heights. This highlights the similarity of the physics after pulses with phases $\varphi$ and $\varphi+2\pi$.
We can furthermore see in \autoref{fig:pulse-interference}(d) that the pulse duration determines the relevance of the memory effects that depend on $e^{i\varphi}$. For short times these effects are dominant~\cite{Gurvitz21}, while for longer pulse durations the mere pulse height or maximum voltage becomes more relevant, following the physical intuition that the memory effects must be limited to some coherence time.

The memory effects which we have discussed for voltage pulses also explain the \enquote{ringing} of the current in \autoref{fig:higher-harmonics}(b) for harmonic driving, $V(t)=\vac\cos(\Omega t)$.
Also in this case electrons hopping between the reservoirs will obtain a phase due to the driving.
For hopping processes that contribute to the current at time $t$ we expect interference with earlier hopping processes that yield a phase $\int_{t^*}^t V(s) ds$ where $t^*$ is the earlier hopping time.
Out of these different phases for different times $t^*$ those will be prominent, for which a whole range of hopping times $t^*$ yields approximately the same phase such that the hopping processes interfere constructively, which happens if $V(t^*)=0$.
Therefore we expect interference effects in the current $I(t)$ that appear as oscillations of the form $\cos(\int_{t^*}^t V(s) ds)$ where $t^*$ is chosen such that $V(t^*)=0$ and $t-t^*$ is small.
These are the oscillations which we observe in \autoref{fig:higher-harmonics}(b) as one can see by plotting $I(t)$ versus $\int_{t^*}^t V(s) ds$ parametrized by $t$ in \autoref{fig:higher-harmonics}(c).
The prominent structures in the current on short time scales in \autoref{fig:higher-harmonics}(b) even for harmonic driving illustrate the relevant of the memory contained in the system-reservoir correlations.

 \section{Summary and Outlook}
\label{sec:summary}

In this paper we extended the real-time renormalization group approach of Refs.~\cite{Pletyukhov12a,Reininghaus14} to periodically driven systems.
Using Floquet theory in Liouville space and the example of the isotropic spin $1/2$ Kondo model we showed how to derive RG equations for such systems at zero temperature in close analogy to the case without periodic driving.
Solving these RG equations numerically with initial conditions that are derived from perturbation theory yields the Fourier series representation of the current and the differential conductance.

For the harmonically driven Kondo model we confirmed that sufficiently fast and strong driving leads to satellite peaks of the Kondo resonance in the time-averaged differential conductance. We have seen that both decoherence due to the periodic driving and photon-assisted tunneling are required for a quantitative description of the differential conductance.
The FRTRG includes both effects by self-consistently including self-energy insertions and shows good quantitative agreement with experimental data.
Using the FRTRG we can compute the time-resolved current in the full crossover from weak to strong driving and find good agreement with predictions for limiting cases.

The micromotion of the current through a strongly and rapidly driven system shows a ringing structure that has been described before for different systems~\cite{Jauho94}. We analyzed the memory in the system-reservoir correlations that leads to these rapid oscillations in the current by studying short, isolated voltage pulses.
For sufficiently short pulses the FRTRG results confirm that the current following after the pulse due to memory effects is mainly characterized by the phase $e^{i\varphi}$ where $\varphi=\int V(t) dt$. This can lead to such counter-intuitive effects as a weaker pulse (with $\varphi=\pi/2$) transporting more charges than a stronger pulse (with $\varphi=3\pi/2$).

The $E$-flow scheme of the real-time renormalization group has been applied to different open quantum systems in nonequilibrium at zero and finite temperature \cite{Pletyukhov12a,Kashuba13a,Reininghaus14,Horig14,Lindner19}.
Also the transient and quench dynamics have been studied using this RG method~\cite{Kashuba13b,Kennes13}.
The here developed Floquet theory applied to the real-time RG allows us to study many other  periodically driven open systems~\cite{Eissing16,Eissing16b,Mathey20} such as the prototypical spin-boson or interacting resonant level model.
Concerning further directions, a combination of periodic driving and finite temperature in the real-time RG seems achievable using RG equations similar to those considered in Ref.~\cite{Reininghaus14}, albeit this will require a higher numerical effort.
The ability to study time-dependent systems makes it possible to apply the real-time renormalization group also to quantum dots with superconducting reservoirs at finite bias voltage.
Also an extension to multilevel systems as done in Ref.~\cite{Lindner19} seems feasible.
The successful application to coherently driven systems highlights the flexibility of the real-time RG for nonequilibrium quantum systems with strong correlations.

\begin{acknowledgments}
We thank A.~Kogan for providing the experimental data and for very useful discussions.
This work was supported by the Deutsche Forschungsgemeinschaft (DFG, German Research Foundation) via RTG 1995 and Germany's Excellence Strategy -- Cluster of Excellence Matter and Light for Quantum Computing (ML4Q) EXC 2004/1 -- 390534769.
We acknowledge support from the Max Planck-New York City Center for Non-Equilibrium Quantum Phenomena.
\end{acknowledgments}

\appendix
\section{Propagator in Liouville-Floquet space}
\label{app:floquet}
In this appendix we derive some rules for Floquet matrices which lead to \Eq{eq:pi-floquet-space}.
The basic property of Floquet matrices is that convolutions in time domain become matrix products in Floquet space. Consider two functions $a(t,t')$ and $b(t,t')$ of two time arguments and denote by $A$ and $B$ the corresponding Floquet matrices. We define $C\coloneqq AB$ and prove below that the corresponding function in time domain is
\begin{equation}
  c(t,t') = \int_{t'}^t ds\,a(t,s)\,b(s,t') \eqqcolon [a*b](t,t') \,.
\end{equation}
The definition of Floquet matrices in Eqs.~\eqref{eq:floquet_1} and \eqref{eq:floquet-matrix} can equivalently be written as
\begin{gather}
  A(E)_{nm} = \int_0^\infty ds\,e^{i(E+n\Omega)s} \int_0^T \frac{dt}{T} e^{i(n-m)\Omega t} a(t+s, t)\,, \\
  a(t+s, t) = \int_{-\infty}^\infty \frac{dE}{2\pi}\, e^{-i(E+n\Omega)s} \sum_{n\in\mathbb{Z}} e^{-in\Omega t} A_{n0}(E)\,.
\end{gather}
The product of two Floquet matrices expressed in time domain is
\begin{widetext}
\begin{align}
C(E)_{nm}
  &= \sum_k \iint_0^\infty ds\,ds'\,e^{i(E+n\Omega)s + i(E+k\Omega)s'} \frac1{T^2} \int_0^T dt\, \int_0^T dt'\, e^{i(n-k)\Omega t + i(k-m)\Omega t'} a(t+s, t) b(t'+s', t') \\
  &= \iint_0^\infty ds\,ds'\,e^{i(E+n\Omega)s + iEs'} \frac1T \int_0^T dt\, \int_0^T dt'\, e^{in\Omega t - im\Omega t'} \sum_k \delta(s'+t'-t + kT) a(t+s, t) b(t'+s', t') \\
  &= \iint_0^\infty ds\,ds'\,e^{i(E+n\Omega)s + iEs'} \frac1T \int_0^T dt'\, e^{in\Omega (t'+s') - im\Omega t'} a(t'+s'+s, t'+s') b(t'+s', t') \\
&= \int_0^\infty ds''\,e^{i(E+n\Omega)s''} \frac1T \int_0^T dt'\, e^{i(n-m)\Omega t} \int_0^{s''} ds'\, a(t'+s'', t'+s') b(t'+s', t') \\
  &= \int_0^\infty ds\,e^{i(E+n\Omega)s} \frac1T \int_0^T dt\, e^{i(n-m)\Omega t} [a*b](t+s, t)
\,.
\end{align}
\end{widetext}

We can now use the convolution property to calculate the Floquet matrix of a time-ordered integral. Consider a function $f(t)$ of a single time argument which does not necessarily commute with itself at different time arguments, $[f(t), f(s)]\neq0$. To use the Floquet matrix formalism we define a function of two time arguments as $\tilde{f}(t,t')\equiv f(t')$ and an identity function $o(t,t')\equiv1$. Then we can rewrite a time-ordered integral (skipping time arguments $(t,t')$ on the right hand side) as
\begin{equation}
  \mathcal{T} e^{\int_{t_0}^t ds\, f(s)}
  = o + \tilde{f}*o + \tilde{f}*\tilde{f}*o + \ldots
  = \sum_{k=0}^\infty \tilde{f}^{*k}*o \,.
\end{equation}
The Floquet matrices for $o$ and $\tilde{f}$ are
\begin{align}
  O(E)_{nm} &= \frac{i\delta_{nm}}{E+n\Omega} = \Big[\frac{i}{E+\hat{N}\Omega}\Big]_{nm}\,, \\
  \tilde{F}(E)_{nm} &= \frac{i\hat{f}_{n-m}}{E+n\Omega} = \Big[ \frac{i}{E+\hat{N}\Omega} F \Big]_{nm}\,,
\end{align}
where $\hat{f}_n=\frac1T \int_0^T dt\,e^{in\Omega t} f(t)$ are the Fourier coefficients of $f(t)$, and we defined $\hat{N}_{nm}=n\delta_{nm}$ and $F_{nm}\equiv\hat{f}_{n-m}$.
The Floquet matrix representing $\mathcal{T} e^{\int_{t_0}^t ds\, f(s)}$ is thus given by
\begin{align}
  \sum_{k=0}^\infty \Big[\frac{i}{E+\hat{N}\Omega} F \Big]^k \frac{i}{E+\hat{N}\Omega}
  = \frac{i}{E+\hat{N}\Omega - iF} \,.
\end{align}
For the special case of the propagator $f(t)=-iL_V-iL_R(t)$ and $F=-iL_V-i\hat{L}_R$. As usual, scalars in Floquet space ($E$ and $L_V$) are interpreted as being multiplied with the identity matrix.
This explains the step from \Eq{eq:pi-time-domain} to \Eq{eq:pi-floquet-space}.

\section{Wick's theorem in Liouville space}
\label{app:wick}

We derive Wick's theorem in Liouville space by starting from the general form
\begin{equation}
  \tr_R \left( J_1^{p_1} J_2^{p_2} \cdots J_n^{p_n} \rho_R^\mathrm{eq} \right)
  \label{eq:wick-derivation-1}
\end{equation}
where $n$ is even.
Our aim is to reduce this to the known form of Wick's theorem for creation and annihilation operators. This can be achieved by reordering the superoperators and turning $J^-$ to $J^+$ operators. Once we have an expression of the form in \Eq{eq:wick-derivation-1} with all $p_i=+$ we can simply replace all superoperators $J^+_i$ by operators $c_i$.

We start manipulating \Eq{eq:wick-derivation-1} by setting $p_1=+$. This does not change the result of the expression as one can conclude from the definition of $J_1^p$ since the trace is cyclic.
Next we consider $J_2^{p_2}$ in \Eq{eq:wick-derivation-1}. If $p_2=+$ we do not change anything. For $p_2=-$ we can use that $J_2^{p_2}$ is the last superoperator (counted from the right) with Keldysh index $-1$. Thus this superoperator denotes right multiplication of the whole expression inside the trace by $c_2$. As the trace is cyclic, we can instead multiply with $c_2$ from the left and obtain
\begin{subequations}\begin{align}
      \tr_R \left( J_2^+ J_1^+ \cdots J_n^{p_n} \rho_R^\mathrm{eq} \right) \text{ if }p_2=- \,. \\
      \tr_R \left( J_1^+ J_2^+ \cdots J_n^{p_n} \rho_R^\mathrm{eq} \right) \text{ if }p_2=+ \,.
    \end{align}\label{eq:wick-derivation-2}\end{subequations}
This expression is still equivalent to \Eq{eq:wick-derivation-1}.
By iterating the last step we obtain an expression with reordered superoperators that all have Keldysh index $+1$.
The new order of the superoperators is obtained by first separating superoperators with Keldysh indices $+1$ and $-1$, e.g., $J_1^+ J_2^- J_3^- J_4^+ J_5^- J_6^+ \to J_2^- J_3^- J_5^- J_1^+ J_4^+ J_6^+$, and then reverting the order of the superoperators with Keldysh index $-1$: $J_2^- J_3^- J_5^- J_1^+ J_4^+ J_6^+\to J_5^- J_3^- J_2^- J_1^+ J_4^+ J_6^+$.
Regarding the example, we have just proven
\begin{equation}
  \tr_R \left( J_1^+ J_2^- J_3^- J_4^+ J_5^- J_6^+ \rho_R^\mathrm{eq} \right)
  = \tr_R \left( c_5 c_3 c_2 c_1 c_4 c_6 \rho_R^\mathrm{eq} \right)
  \,.
  \label{eq:wick-derivation-3}
\end{equation}

We can now use Wick's theorem to evaluate the right hand side of \Eq{eq:wick-derivation-3}:
\begin{align}
  \left< c_1 c_2 \cdots c_n \right>
  &= \sum_P (-1)^P \left< c_{P_1} c_{P_2} \right> \cdots \left< c_{P_{n-1}} c_{P_n} \right> \\
  \left< c_1 c_2 \right>
  &= \tr_R \left( c_1 c_2 \rho_R^\mathrm{eq} \right)
  = \delta_{1\bar2} \frac{D(\omega_1)}{1+e^{\eta_1\beta\omega_1}}
\end{align}
where $\beta$ denotes inverse temperature and $P$ runs over all possible combinations of pairs of indices like $(P_1,P_2)$ where within each pair the order of the indices is preserved (i.e. $P_1<P_2$). $(-1)^P$ adds a sign when the total number of permutations of indices in $P$ is odd.

Before we can use Wick's theorem for superoperators we need to analyze the effect of the reordering described above \Eq{eq:wick-derivation-3}.
Let us consider one specific choice of contractions or one specific permutation $P$. Writing the contractions as lines we obtain a minus sign for each crossing of contraction lines.
\begin{equation}
  \tr_R \big( \wick{\c1J_1^+ \c2J_2^- \c1J_3^- \c3J_4^+ \c3J_5^- \c2J_6^+} \rho_R^\mathrm{eq} \big)
  = \tr_R \big(\wick{\c3c_5 \c1c_3 \c2c_2 \c1c_1 \c3c_4 \c2c_6} \rho_R^\mathrm{eq} \big)
  \,.
  \label{eq:wick-derivation-4}
\end{equation}
When changing the order of the superoperators we also change the number of line crossings. For each superoperator $J_i^-$ with the contraction line directed to the right (before reordering) the number of contraction line crossings changes by an even number if $i$ is odd and by an odd number if $i$ is even.
When the contraction line of $J_i^-$ is directed to the left, the number of contraction line crossings changes by an even number if $i$ is even and by an odd number if $i$ is odd.
Thus during the whole procedure we obtain an extra sign beyond the sign from the number of contractions of
\begin{equation}
  \Big(\Pi_{i\text{ even}} p_i\Big) \Big(\Pi_{i\text{ even}} p_{P_i}\Big) \,.
\end{equation}
Here $P_i$ for even $i$ selects the index of the right superoperator of each contraction line.

Furthermore in all contraction pairs with a $J^-$ standing on the right, the order of the superoperators within the contraction pair has been inverted while reordering the superoperators. This can be compensated by using
\begin{equation}
  \left<c_1c_2\right>
  = e^{-\beta\omega_1\eta_1} \left<c_2c_1\right>.
\end{equation}

By putting these considerations together we obtain
\begin{align}
  &\tr_R \Big( p_2 J_1^{p_1} J_2^{p_2} \cdots p_n J_{n-1}^{p_{n-1}} J_n^{p_n} \rho_R^\mathrm{eq} \Big) \notag \\
  &= \sum_P (-1)^P p_{P_2} \left< J_{P_1}^{p_{P_1}}J^{p_{P_2}}_{P_2}\right> \cdots p_{P_n} \left< J_{P_{n-1}}^{p_{P_{n-1}}}J^{p_{P_n}}_{P_n}\right>
\end{align}
with the contraction lines defined by
\begin{equation}
  p_2 \left<J_1^{p_1} J_2^{p_2}\right>
  = p_2 \delta_{1\bar2} \frac{D(\omega_1)}{1+e^{p_2\eta_1\beta\omega_1}}
  = \gamma^{p_1p_2}_{12} \,.
\end{equation}

\section{Notation}
\label{app:notation}
An overview of the variables used in the derivation of the RG equations can be found in \autoref{tab:notation}.
\begin{table}[bt]
\caption{\label{tab:notation}
Variables used in the derivation of the method.
}
\begin{tabular}{p{.19\linewidth} p{.59\linewidth} p{.17\linewidth}}
    \toprule
    Symbol & Description & Reference \\
    \midrule
${J_{\alpha\alpha'}^{(0)}=2\sqrt{x_\alpha x_{\alpha'}}J^{(0)}}$ & \qquad\qquad bare coupling & \autoref{sec:model} \\
$\mu_\alpha(t), \hat\mu_\alpha$ & chemical potential of reservoir $\alpha$ & \eqref{eq:reservoir-liouvillian}, \eqref{eq:L_R-floquet-space} \\
$D, D(\omega)$ & band width, cutoff function & \autoref{sec:model} \\
$\Omega=2\pi/T$ & driving frequency & \\
    ${\hat{N}_{nm}=n\delta_{nm}}$ & \qquad diagonal Floquet matrix & \\
$\Leff(E)$ & effective Liouvillian & \eqref{eq:L-floquet-space} \\
    $L_R(t)$, $\hat{L}_R$ & reservoir Liouvillian & \eqref{eq:reservoir-liouvillian},~\eqref{eq:L_R-floquet-space} \\
    $J^p_1$ & fermionic superoperators & \eqref{eq:fermionic-superoperators} \\
    $V, L_V$ & coupling Hamiltonian, Liouvillian & \eqref{eq:coupling-hamiltonian}, \eqref{eq:vertex-liouvillian} \\
$G_{11'}^{(0)pp'}$ & bare coupling vertex & \eqref{eq:vertex-liouvillian} \\
$\hat{R}^{(0)}_{12}$ & bare resolvent & \eqref{eq:bare-resolvent} \\
    $\bar\omega_X, \hat{\bar\mu}_X$ & $\bar\omega_{1\ldots}=\eta_1\omega_1+\ldots$, analogous for $\hat{\bar\mu}_X$ & below~\eqref{eq:bare-resolvent} \\
$\gamma^{pp'}_{11'}(\omega,\omega'), \gamma^p(\omega)$ & \qquad\qquad reservoir contractions & \eqref{eq:reservoir-contraction} \\
$\Gkavg_{11'}$ & vertex summed over Keldysh indices & below~\eqref{eq:L-leading-order-simplified} \\
$\hat{R}(E),\hat{R}_X(E)$ & \qquad effective resolvent & \eqref{eq:effective-resolvent}, \eqref{eq:effective-resolvent-argument} \\
    $\Gfull_{11'}^{pp'}(E;\bar\omega,\bar\omega')$ & \qquad full effective vertex & \eqref{eq:G-diagrams-1} \\
$\Gfull_{11'}^{pp'}(E)$ & effective vertex, $\bar\omega=\bar\omega'=0$ & \eqref{eq:effective_vertex} \\
${\hat{E}_{12} = E - \hat{\bar\mu}_X}$ & \qquad shifted energy & above~\eqref{eq:resolvent-frequency-approx-full} \\
    $\chi(E),Z(E)$ & used in approximation of $R(E)$ & \eqref{eq:resolvent-frequency-approx},  \eqref{eq:Z-definition} \\
$\Delta$ & energy scale of bias voltage & below~\eqref{eq:G-integral-2} \\
    $\tilde{E}$ & renormalized energy scale & below~\eqref{eq:G-integral-2} \\
$\Gnoeta_{11'}(E)$ & $\Gkavg_{11'}(E)$ with $\eta=+,\eta'=-$ & above~\eqref{eq:L-RGeq-general} \\
$I^\gamma$ & current in lead $\gamma$ & above~\eqref{eq:current-vertex} \\
    $\Ifull_{11'}^{pp'},\Inoeta_{11'}$ & current vertex, analogous to $\Gnoeta_{11'}$ & \eqref{eq:current-vertex} \\
    $\Sigma_\gamma$ & current kernel & \eqref{eq:observable-kernel-time-domain} \\
$\delta\hat{\bar\mu}_{12}\propto\id$ & infinitesimal variation of $\hat{\bar\mu}_{12}$ & \\
    $\delta L(E),\delta R_X(E)$ & \qquad variation of $L(E)$ and $R_X(E)$ & \eqref{eq:variation-resolvent} \\
$\Gamma(E)$ & spin relaxation rate & \eqref{eq:L-parametrization} \\
    $\Gamma^\gamma(E)$ & parametrization of $\Sigma_\gamma(E)$ & \eqref{eq:current-kernel-parametrization} \\
    $I^\gamma_{11'}(E)$ & parametrization of $\Inoeta^\gamma_{11'}(E)$ & \eqref{eq:I-parametrization} \\
    $L^{a,b}$ & superoperator algebra & \eqref{eq:La}, \eqref{eq:Lb} \\
    $\hat{L}^{1,2,3,a,b}_{\sigma\sigma'}$ & superoperator algebra & \eqref{eq:La-hat}--\eqref{eq:L3-hat} \\
    $G^{a,2,3}_{11'}(E)$ & parametrization of $\Gnoeta(E)$ & \eqref{eq:G-parametrization} \\
$\delta\Gamma(E),\delta\Gamma^\gamma(E)$ & \qquad variation of $\Gamma(E),\Gamma^\gamma(E)$ & \\
${\Lambda=\im E,\Lambda_0}$ & \qquad\mbox{RG flow parameter, starting value} & \\
$J,K,M$ & $G^2_{12}, G^3_{12}, I^\gamma_{12}$ in equilibrium RG flow & \mbox{\eqref{eq:G2-parametrization-J}--\eqref{eq:I-parametrization-M}} \\
\bottomrule
\end{tabular}
\end{table} 

\section{Frequency integral}
\label{app:integral}

In this appendix we solve the integral in \Eq{eq:G-integral-1}.
With simplified notation we want to solve
\begin{equation}
  I \coloneqq \int_0^\infty d\omega\, \frac1{\omega+\hat\chi_1} \hat{A} \frac1{\omega+\hat\chi_2}
\end{equation}
where $\hat\chi_1$, $\hat\chi_2$ and $\hat{A}=Z(E)G_{12}^{p_1p_2}(E)$ are Floquet matrices.
For the approximation which we will use for a more efficient numerical evaluation of the result, we need to use some properties of $\chi_1$, $\chi_2$ and $\hat{A}$.
First of all, we can ignore here that \Eq{eq:G-integral-1} contains superoperators in Liouville spaces because all superoperators will be parametrized by plain Floquet matrices which appear as scalars in Liouville space.
Second, we use that $G^{p_1p_2}_{12}(E)$ is in leading order a diagonal Floquet matrix. All off-diagonal contributions to the effective vertex come from diagrams of order $O(G^2)$.
Similarly, also in $Z(E)$ and $\chi(E)$ the diagonal is dominant and all off-diagonal matrix elements are small compared to the diagonal.
Assuming that we can diagonalize $\hat\chi_i=\sum_k \ket{k^i}\chi^i_k \bra{k^i}$, we can solve the integral analytically:
\begin{align}
  &I = \sum_{kl} \ket{k^1}\!\bra{k^1}\hat{A}\ket{l^2}\!\bra{l^2} \int_0^\infty d\omega\, \frac1{\omega+\chi^1_k} \frac1{\omega+\chi^2_l} \\
  &= \sum_{kl} \ket{k^1}\!\bra{k^1}\hat{A}\ket{l^2}\!\bra{l^2} \frac{1}{\chi^1_k - \chi^2_l} \log\Big(\frac{\chi^1_k}{\chi^2_l}\Big) \\
  &= \sum_{kl} \ket{k^1}\!\bra{k^1}\hat{A}\ket{l^2}\!\bra{l^2} \frac{1}{2} \Big[ \frac1{\chi^1_k} + \frac1{\chi^2_l} + O\big(\tfrac{(\chi_k^1-\chi_l^2)^2}{(\chi_k^1+\chi_l^2)^3}\big) \Big] \\
  &= \frac{1}{2} \Big[ \frac1{\hat\chi_1} \hat{A} + \hat{A} \frac1{\hat\chi_2} \Big] + O(\Delta)\,.
\end{align}
Since $\hat{A}$ is approximately diagonal and off-diagonal matrix elements fall off quickly, and since also $\hat\chi_1\approx\hat\chi_2$ is approximately diagonal, the combination ${\bra{k^1}\hat{A}\ket{l^2}}{(\chi_k^1-\chi_l^2)^2}$ will always be small. At vanishing driving ($\Delta=0$), all matrices are diagonal, $\hat\chi_1=\hat\chi_2$, and the approximation is exact. This lets us conclude that the deviation must be at least of order $O(\Delta)$.
Inserting now $\omega=-p_5\bar\omega_4$, $\hat\chi_1=p_5\chi(\hat{E}_{34})$, $\hat\chi_2=p_5\chi(\hat{E}_{1234})$ and $\hat{A}=Z(\hat{E}_{34})G_{12}^{p_1p_2}(\hat{E}_{34})$ we obtain \Eq{eq:G-integral-2}.

\section{Liouville space parametrization and symmetry}
\label{app:symmetries}
In this appendix we derive the parametrizations for the effective Liouvillian and vertex.
A rotation by a vector $\mathbf{r}$ is represented in spin space by a unitary matrix $U=\exp(-i\mathbf{r}\cdot\boldsymbol{\sigma})$.
The rotation of density matrices is described by a superoperator $\mathcal{U}\bullet=U\bullet U^\dag$.
We will use the decomposition of superoperators in the basis of Pauli matrices $\{\sigma_i\}$, $\sigma_0=\id$, with the notation $\Ket{a}\Bra{b}\coloneqq a \tr(b^\dag \bullet)$.

\paragraph{Liouvillian.}
The Liouvillian $L$ is invariant under a rotation if $\mathcal{U}^\dag L\mathcal{U}=L$ or, equivalently, $[\mathcal{U},L]=0$.
This symmetry enforces that $L$ does not contain any terms of the form $\Ket{\sigma_i}\Bra{\sigma_j}$ where $i\neq j$:
In this case we can always find a rotation $U$ such that $U\sigma_iU^\dag=-\sigma_i$ and $U\sigma_jU^\dag = +\sigma_j$ or vice versa. For $i\neq0$ and $j\neq0$ one would choose $U=\sigma_j$, otherwise $U=\sigma_j$ for $i=0$ and $U=\sigma_i$ for $j=0$.
This rotation in spin space yields $\mathcal{U}\Ket{\sigma_i}\Bra{\sigma_j}=-\Ket{\sigma_i}\Bra{\sigma_j}\mathcal{U}$. Thus no terms of the form $\Ket{\sigma_i}\Bra{\sigma_j}$, $i\neq j$ may appear in the decomposition of $L$.
The only remaining terms are $\Ket{\id}\Bra{\id}$, which has to vanish because $L$ must be traceless, and $\sum_{i=1}^3 \Ket{\sigma_i}\Bra{\sigma_i}$. In this sum all terms need to have the same prefactor again due to the symmetry.
This completes the proof that the $L\propto L^a=\frac12 \sum_{i=1}^3 \Ket{\sigma_i}\Bra{\sigma_i}$.

\paragraph{Interaction vertex.}
When parametrizing the effective vertex $\Gnoeta_{11'}$ we additionally need to take into account the spin indices $\sigma,\sigma'$ in the multi-indices $1,1'$. These reservoir-space spin indices can be included in the rotation by expanding also the $\sigma,\sigma'$ matrix space in terms of Pauli matrices:
\begin{equation}
  \Gnoeta_{11'} = \sum_{m=0}^3 G^m_{\alpha\alpha'} \sigma^m_{\sigma\sigma'} \,.
\end{equation}
Now rotational invariance of $G$ can be written in the form
\begin{equation}
  \sum_{m=0}^3 (\mathcal{U} G^m_{\alpha\alpha'} \mathcal{U}^\dag) \, (U\sigma^mU^\dag)_{\sigma\sigma'}
  = \Gnoeta_{11'} \,.
\end{equation}
A further expansion of $G$ is of the form (for simplicity we do not write the indices $\alpha,\alpha'$)
\begin{equation}
  \sum_{m,i,j=0}^3 g^m_{ij} \Ket{\sigma_i}\Bra{\sigma_j} \sigma^m_{\sigma\sigma'} \,.
\end{equation}
A rotation maps this to
\begin{equation}
  \sum_{m,i,j=0}^3 g^m_{ij} \Ket{U\sigma_iU^\dag}\Bra{U\sigma_jU^\dag} (U\sigma^mU^\dag)_{\sigma\sigma'} \,.
\end{equation}

For $m=0$ we find in analogy to the effective Liouvillian that only coefficients with $i=j$ may be nonzero. Knowing that all coefficients with $i=0$ also need to vanish (these terms have nonzero trace), we see that the most general form for $m=0$ is
\begin{equation}
  g^0_{ij}=\tfrac12 G^a \delta_{ij}(1-\delta_{i0}) \,.
\end{equation}
For $m\neq0$ we can use rotations by $U=\sigma_m$, $\sigma_j$, and $\sigma_i$ to show by a similar argument that the coefficients $g^k_{ij}$ can only be nonzero if $m=i$ and $j=0$, or if $\{i,j,k\}$ is a permutation of $\{1,2,3\}$.
The case $j=0$ leads to coefficients of the form
\begin{equation}
  g^m_{i0}=\tfrac12 G^3 \delta_{im}(1-\delta_{i0}) \,.
\end{equation}
For the case where all indices are different and nonzero, a rotation by $U=\frac12(\id+iX+iY+iZ)$ can be used to show that $g^m_{ij}$ is invariant under an equal shift of all indices, e.g. $g^1_{23}=g^2_{31}=g^3_{12}$.
Furthermore, using a rotation by $U=\frac1{\sqrt2}(\id+i\sigma_k)$ we can show that $g^k_{ij}=-g^k_{ji}$.
This leads to a parametrization of the form
\begin{equation}
  g^k_{ij}=i\tfrac14 \varepsilon_{ijk}G^2 \quad\text{ for } i,j,k\in\{1,2,3\}\,.
\end{equation}
This shows that the three $2\times2$ matrices $G^a_{\alpha\alpha'},G^2_{\alpha\alpha'},G^3_{\alpha\alpha'}$ are sufficient to parametrize $\Gnoeta_{11'}$.

\paragraph{Current vertex.}
The current vertex comes with the simplification that the output states can be restricted to $\Ket{\id}$ since we are only interested in the trace of the current vertex. Furthermore, the current vertex only acts on traceless states.
This leads to the general parametrization
\begin{equation}
  \Inoeta^\gamma_{11'} = \sum_{m=0}^3 \sum_{j=1}^3 i_j^m \Ket\id \Bra{\sigma_j} \sigma^m_{\sigma\sigma'}
\end{equation}
for the current vertex, for which the rotational symmetry can be written in the form
\begin{multline}
  \sum_{m=0}^3 \sum_{j=1}^3 i^m_j \Ket\id \Bra{U\sigma_j U^\dag} \otimes (U\sigma^mU^\dag) \\
  = \sum_{m=0}^3 \sum_{j=1}^3 i^m_j \Ket\id \Bra{\sigma_j} \otimes \sigma^m\,.
\end{multline}
Inserting the Pauli matrices for $U$ shows that the rotational symmetry requires $i^m_j\propto\delta_{mj}$.

\section{Superoperator algebra}
\label{app:superoperator-algebra}
In this appendix we derive Eqs.~\eqref{eq:superoperator-algebra-1}, \eqref{eq:superoperator-algebra-2} and \eqref{eq:superoperator-algebra-3}. Implicitly summing over all multiply appearing indices we find for the RG equation for $G^2$:
\begin{align}
  &\hat{L}^2_{\sigma_3\sigma_4} \hat{L}^2_{\sigma_1\sigma_2} \hat{L}^2_{\sigma_4\sigma_3} \\
&= -\frac1{16} \varepsilon_{ijk} \varepsilon_{lmn} \tr_\sigma(\sigma^k\sigma^n) \Ket{\sigma_i} \Bra{\sigma_j} \hat{L}^2_{\sigma_1\sigma_2} \Ket{\sigma_l}\Bra{\sigma_m} \\
  &= -\frac{i}{8} \varepsilon_{ijk} \varepsilon_{lmk} \varepsilon_{jlp} \Ket{\sigma_i} \sigma^p_{\sigma_1\sigma_2} \Bra{\sigma_m} \\
  &= -\frac{i}{8} \varepsilon_{jip} \Ket{\sigma_i} \sigma^p_{\sigma_1\sigma_2} \Bra{\sigma_j}
  \,=\, \frac12 \hat{L}^2_{\sigma_1\sigma_2} \,.
\end{align}
For \Eq{eq:superoperator-algebra-2} and the RG equation for $G^3$ we compute:
\begin{align}
  &\hat{L}^2_{\sigma_3\sigma_4} \hat{L}^2_{\sigma_1\sigma_2} \hat{L}^3_{\sigma_4\sigma_3} \\
  &= \frac{i}{8} \varepsilon_{ijk} \tr_\sigma(\sigma^k\sigma^l) \Ket{\sigma_i} \Bra{\sigma_j} \hat{L}^2_{\sigma_1\sigma_2} \Ket{\sigma_l}\Bra{\id} \\
  &= -\frac14 \varepsilon_{ijk} \varepsilon_{jkm} \Ket{\sigma_i} \sigma^m_{\sigma_1\sigma_2} \Bra{\id} \\
  &= -\frac12 \Ket{\sigma_m} \sigma^m_{\sigma_1\sigma_2} \Bra{\id}
  \,=\, - \hat{L}^3_{\sigma_1\sigma_2} \,.
\end{align}
For \Eq{eq:superoperator-algebra-3} and the RG equation for $I^\gamma$ we use:
\begin{align}
  &\hat{L}^1_{\sigma_3\sigma_4} \hat{L}^2_{\sigma_1\sigma_2} \hat{L}^2_{\sigma_4\sigma_3} \\
  &= -\frac18 \tr_\sigma(\sigma^i \sigma^n) \varepsilon_{ijk} \varepsilon_{jmn}\sigma^k_{\sigma_1\sigma_2} \Ket\id \Bra{\sigma_m} \\
  &= -\frac12 \sigma^k_{\sigma_1\sigma_2} \Ket\id \Bra{\sigma_k}
  \,=\, - \hat{L}^1_{\sigma_3\sigma_4} \,.
\end{align}

\section{Convergence checks}
\label{app:g-self-consistency}

\begin{figure}[t]
  \centering
  \includegraphics[width=\linewidth]{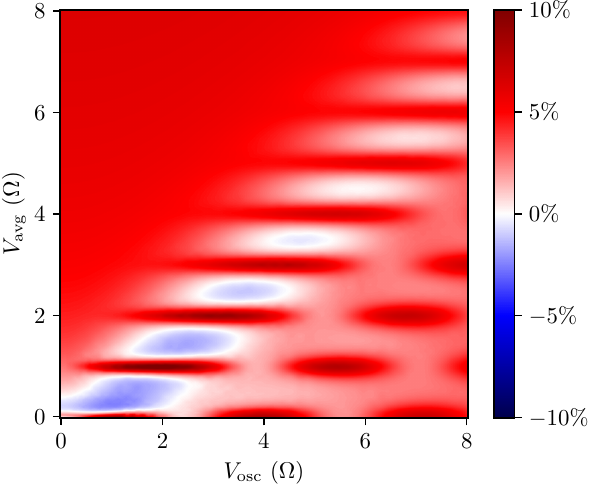}
  \caption{\label{fig:order-comparison}
    Relative difference $(\gdc-\gdc^\text{leading order})/\gdc$ of the differential conductance for $\Omega=7.55\tkv$.
    Here $\gdc^\text{leading order}$ is the differential conductance from an FRTRG calculation that keeps only the leading order in $J$. For all other data we include the next-to-leading order in $J$.
    The dark red spots are the peaks in the $G$ which are overestimated in their height by $\gdc^\text{leading order}$.
    The maximum deviation for these parameters is approximately $10\%$ in the regime where the energy scales $\vac$ and $\Omega$ are approximately equal.
  }
\end{figure}
\begin{figure}[t]
  \centering
  \includegraphics[width=\linewidth]{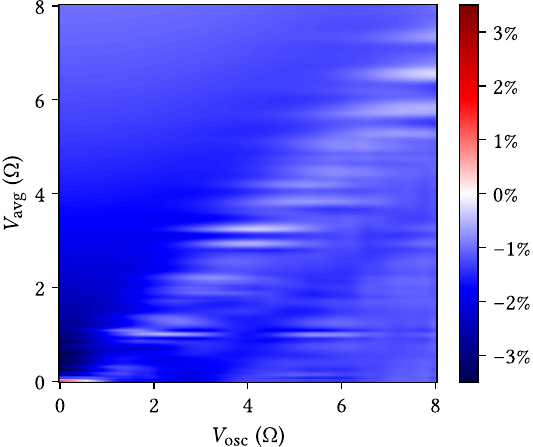}
  \caption{\label{fig:g-self-consistency}
    Relative difference $(\gdc^{\delta\Gamma^\gamma}-\gdc^{\Gamma^\gamma})/\gdc^{\delta\Gamma^\gamma}$ of the differential conductance $\gdc^{\delta\Gamma^\gamma}=\delta\Gamma^\gamma(0)_{00}$ and the numerical derivative of the current $\gdc^{\Gamma^\gamma}=d\Gamma^\gamma(0)_{00}/d\vdc$ for $\Omega=7.55\tkv$.
    The maximum deviation for these parameters is approximately $3.5\%$ and $\gdc^{\Gamma^\gamma}$ is in general slightly larger than $\gdc^{\delta\Gamma^\gamma}$ except near equilibrium.
    The values for $\gdc^{\delta\Gamma^\gamma}$ used in this comparison are shown in \autoref{fig:G-vdc-vac}.
  }
\end{figure}

In the derivation of the RG equations we use an expansion in the coupling $J$ up to next-to-leading order.
To check the convergence of this approximation we compare our results to an FRTRG calculation which keeps only the leading order in $J$~\cite{Pletyukhov12a}. This comparison is shown in \autoref{fig:order-comparison} for the same parameters as in \autoref{fig:G-vdc-vac}.
One can see that the effect of the next-to-leading order on the differential conductance $G$ is at most $10\%$ for the chosen parameters.
In general the FRTRG in leading order overestimates the satellite peaks in the differential conductance. This is precisely the effect which we expect when decoherence of the quantum state is underestimated because some processes contributing to decoherence are neglected in the leading order approximation.

As an additional check we have included in the RG equations the current and the differential conductance like independent variables. To test the accuracy and self-consistency of our method we compare the differential conductance that is directly included in the RG equations ($\delta\Gamma^\gamma$) with the numerical derivative of the current ($\Gamma^\gamma$).
An overview of this comparison for the same parameters as in \autoref{fig:G-vdc-vac} is shown in \autoref{fig:g-self-consistency}.
The good agreement of both ways of computing the conductance indicates that the contribution of higher order terms in the RG equations for $\delta\Gamma^\gamma$ and $\Gamma^\gamma$ is not significant. However, since both ways of calculating the conductance involve the same current vertex and other renormalized quantities, this comparison does not prove the validity of the full RG equations.

\begin{figure}[t!]
  \centering
  \includegraphics[width=\linewidth]{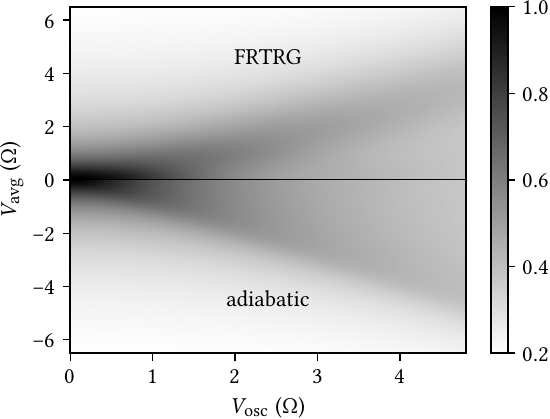}
\caption{\label{fig:calibration-problems}Differential conductance $\gdc$ in units of $2e^2/h$ as calculated using FRTRG for $\Omega=0.75\tkv$ (upper panel) and its adiabatic approximation (lower panel).
    The comparison illustrates that in the full FRTRG results the maximum in $\gdc$ is shifted slightly towards smaller $\vdc$.
    Thus a calibration of $\vac$ by comparison to the adiabatic limit will be biased and underestimate $\vac$.
    We stress that this does not simulate the precise Kondo temperature used for the calibration in Ref.~\cite{Kogan04}, but only aims for a qualitative comparison with the calibration measurement shown in Fig.~2 of Ref.~\cite{Kogan04}.
  }
\end{figure}

\section{Calibration mismatch in comparison to the experiment of Ref.~\cite{Kogan04}}
\label{app:experiment}

In \autoref{sec:experiment} and \autoref{fig:experiment} the comparison of the FRTRG data to a measurement of a single electron transistor in the Kondo regime~\cite{Kogan04} is explained.
The adjustment of the calibration of $\vac$---unavoidable for finding an agreement between our results and the experiment---is justified by the calibration procedure used in the experiment.
In \autoref{fig:calibration-problems} the differential conductance and its adiabatic limit are shown in a similar setting as what was used to calibrate $\vac$ in Ref.~\cite{Kogan04} (Fig.~2 there).
We qualitatively find that calibrating $\vac$ by comparison to a measurement of the differential conductance to a simulation of its adiabatic limit will underestimate $\vac$.
However, we note that the factor $1.4$ by which we need to adjust $\vac$ to find good agreement between theory and experiment in \autoref{fig:experiment} is larger than what this qualitative simulation of the calibration procedure suggests.

 {}

\end{document}